\documentclass[aps,prl,twocolumn,longbibliography,superscriptaddress,notitlepage,floatfix]{revtex4-2}
\usepackage{amsmath,amsfonts,amssymb,color,epsfig,graphics,graphicx,latexsym,theorem,url,multirow,diagbox}
\usepackage[colorlinks,citecolor=blue,linkcolor=blue,urlcolor=blue]{hyperref}
\usepackage[utf8]{inputenc}
\usepackage{booktabs}
\usepackage[T1]{fontenc}
\usepackage{courier}
\usepackage{listings}
\usepackage{latexsym}
\usepackage{balance}
\usepackage{dcolumn}
\usepackage{comment}
\usepackage{times} 
\usepackage{ulem}
\usepackage{xcolor}
\usepackage{color}
\usepackage{pifont} 
\usepackage{newfloat}
\usepackage{placeins}
\usepackage{soul}
\usepackage{braket}
\usepackage{bm}

\begin{document}

\title{A superconducting surface-code processor with lattice-surgery logical operations}

\newcommand{\zju}{School of Physics, ZJU-Hangzhou Global Scientific and Technological Innovation Center, \\and Zhejiang Key Laboratory of Micro-nano Quantum Chips and Quantum Control, Zhejiang University, Hangzhou, China}
\newcommand{\gscaep}{Graduate School of China Academy of Engineering Physics, Beijing, China}

\author{Yanzhe Wang}
\thanks{These authors contributed equally.}
\affiliation{\zju}

\author{Fanhao Shen}
\thanks{These authors contributed equally.}
\affiliation{\zju}

\author{Haipeng Xie}
\affiliation{\gscaep}

\author{Aosai Zhang} 
\affiliation{\zju}
\author{Yu Gao}
\affiliation{\zju}

\author{Chuanyu Zhang}
\affiliation{\zju}
\author{Xuhao Zhu}
\affiliation{\zju}
\author{Feitong Jin}
\affiliation{\zju}
\author{Yiren Zou} 
\affiliation{\zju}
\author{Ning Wang}
\affiliation{\zju}
\author{Zhengyi Cui}
\affiliation{\zju}
\author{Zehang Bao}
\affiliation{\zju}
\author{Zitian Zhu}
\affiliation{\zju}
\author{Jiarun Zhong}
\affiliation{\zju}
\author{Gongyu Liu}
\affiliation{\zju}
\author{Jia-Nan Yang}
\affiliation{\zju}
\author{Yihang Han}
\affiliation{\zju}
\author{Yiyang He}
\affiliation{\zju}
\author{Jiayuan Shen}
\affiliation{\zju}
\author{Han Wang}
\affiliation{\zju}
\author{Jiahua Huang}
\affiliation{\zju}
\author{Xinrong Zhang}
\affiliation{\zju}
\author{Sailang Zhou}
\affiliation{\zju}

\author{Hang Dong}
\affiliation{\zju}
\author{Jinfeng Deng}
\affiliation{\zju}
\author{Yaozu Wu}
\affiliation{\zju}
\author{Zixuan Song}
\affiliation{\zju}

\author{Hekang Li}
\affiliation{\zju}
\author{Zhen Wang}
\affiliation{\zju}
\author{Chao Song}
\affiliation{\zju}
\author{Qiujiang Guo}
\affiliation{\zju}

\author{Pengfei Zhang}
\email{pfzhang@zju.edu.cn}
\affiliation{\zju}

\author{H. Wang}
\email{hhwang@zju.edu.cn}
\affiliation{\zju}

\author{Ying Li}
\email{yli@gscaep.ac.cn}
\affiliation{\gscaep}

\begin{abstract}
Fault-tolerant logical operations are fundamental for scalable quantum computation. Here, we report the experimental realization of lattice-surgery operations between a pair of distance-three surface-code logical qubits on a planar superconducting processor. During repeated syndrome extraction cycles, the logical qubits exhibit per-cycle error rates of $0.0365(2)$ and $0.0282(1)$, respectively, after leakage events are rejected. By leveraging joint initialization and lattice splitting, we deterministically prepare a logical Bell state, confirming genuine bipartite entanglement via the error-corrected logical state fidelity. We further execute a two-qubit Deutsch-Jozsa algorithm at the logical level to demonstrate algorithmic utility in a fault-tolerant framework. Finally, to achieve universal control, we implement magic-state injection and gate teleportation to realize continuous non-Clifford rotations about the logical $X$ axis. For the logical $R_{X}(\pi/4)$ gate, we achieve a logical gate fidelity of $0.943_{-9}^{+10}$ conditioned on the absence of detected errors. These results establish lattice surgery as a practical and versatile paradigm for logical computation in near-term surface-code architectures, representing a critical milestone toward scalable fault-tolerant quantum advantage in superconducting circuits.
\end{abstract}

\maketitle

\section{Introduction}

Quantum error correction (QEC)~\cite{Shor1995Phys.Rev.A., Calderbank1996Phys.Rev.A, Steane1996Phys.Rev.Lett., Knill1997Phys.Rev.A, Terhal2015Rev.Mod.Phys.} is essential for realizing large-scale quantum computation in the presence of noise. Among the diverse range of QEC codes~\cite{Ni2023Nature, Breuckmann2021PRXQuantum, Bravyi2024Nature, Wang2026Nat.Phys., Kitaev2003Ann.Phys.}, the surface code~\cite{Bravyi1998quantum, Dennis2002J.Math.Phys., Kitaev2003Ann.Phys., Fowler2012Phys.Rev.A} is the leading candidate owing to its high fault-tolerance threshold, local stabilizer structure, and compatibility with planar qubit layouts~\cite{Krinner2022Nature, GoogleQuantumAI2023Nature, He2025Phys.Rev.Lett., Acharya2025Nature, Zhang2025Nat.Commun., Ryan-Anderson2024Science, Bluvstein2024Nature, Bluvstein2026Nature}. Over the past several years, experimental progress has enabled the demonstration of repeated syndrome extraction cycles (SECs)~\cite{Krinner2022Nature, GoogleQuantumAI2023Nature, He2025Phys.Rev.Lett.} and the realization of logical memory lifetimes that surpass the break-even point~\cite{Acharya2025Nature}. These advances establish encoded quantum memory as a realistic primitive, shifting the frontier to the next critical challenge: implementing logical operations that preserve fault tolerance while remaining compatible with practical hardware constraints.

Fault-tolerant logical operations implement the desired computation while preventing low-weight physical errors from propagating into uncorrectable logical failures~\cite{Shor1997faulttolerantquantumcomputation}. While transversal operations offer a straightforward approach to fault tolerance, performing them between code blocks necessitates non-local connectivity, which is a feature of reconfigurable systems like trapped ions~\cite{Ryan-Anderson2024Science} and Rydberg atoms~\cite{Bluvstein2024Nature, Bluvstein2026Nature}. For solid-state platforms restricted to static planar layouts, such as superconducting qubits~\cite{Krantz2019Appl.Phys.Rev., Kjaergaard2020Annu.Rev.Condens.MatterPhys.}, lattice surgery has emerged as the foundational paradigm for logical operations. By dynamically deforming surface-code patches, lattice surgery enables various operations among logical qubits under the strict constraints of two-dimensional architectures~\cite{Horsman2012NewJ.Phys., Litinski2019Quantum, Erhard2021Nature, Ryan-Anderson2024Science, Lacroix2025Nature, Besedin2026Nat.Phys., Bluvstein2026Nature}.

Lattice surgery implements computation by measuring joint logical Pauli operators between adjacent logical qubits through the merging and splitting of code patches. These two primitives serve as the building blocks for a universal gate set~\cite{Horsman2012NewJ.Phys., Litinski2019Quantum}, including both Clifford and non-Clifford logical gates. Recent experiments on superconducting quantum processors have demonstrated lattice surgery for logical state teleportation between a pair of distance-three color-code logical qubits~\cite{Lacroix2025Nature} and Bell state preparation by splitting a distance-three surface code into two repetition-code blocks~\cite{Besedin2026Nat.Phys.}. However, the direct demonstration of lattice surgery between fully error-correctable surface-code patches, as well as the realization of non-Clifford logical gates, remains an outstanding experimental challenge.

In this work, we experimentally realize these foundational lattice-surgery operations between two distance-three surface-code logical qubits on a superconducting system of 125 qubits. We utilize this logical processor to investigate fundamental fault-tolerant quantum computation primitives and small-scale programmable logical algorithms. Specifically, by employing a lattice-split operation, we deterministically prepare a Bell state across the two surface-code logical qubits, providing simultaneous protection against bit-flip and phase-flip errors. We then execute a two-qubit Deutsch-Jozsa algorithm~\cite{Gulde2003Nature} at the logical level, utilizing soft post-selection to enhance the classification accuracy. Finally, we implement a magic-state injection and gate-teleportation circuit to achieve continuous non-Clifford logical rotations around the logical $X$ axis, characterizing the resulting logical $R_{X}(\pi/4)$ gate via quantum process tomography. Together, these results provide a comprehensive validation of lattice surgery as a viable route toward scalable logical computation.

\section{Overview of the setup}

\begin{figure*}[tbp]
\centering
\includegraphics[width=\textwidth]{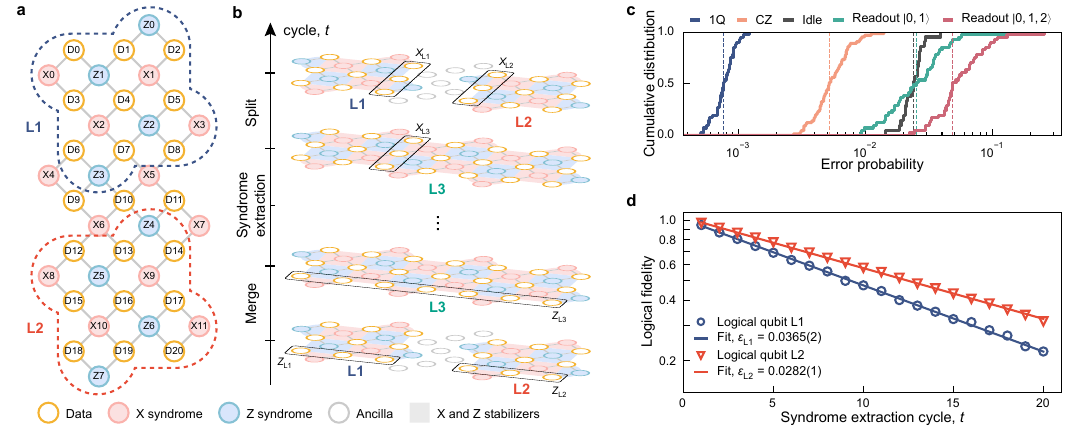}
\caption{
\textbf{Experimental layout and device performance.} 
\textbf{a}, Qubit connectivity used for lattice surgery. The lattice comprises data qubits (gold circles), X-type syndrome qubits (red circles), and Z-type syndrome qubits (blue circles). Each syndrome qubit is coupled to two or four neighboring data qubits, facilitating the stabilizer measurements. 
\textbf{b}, Conceptual illustration of lattice-merge and lattice-split operations. Two distance-three surface-code patches, denoted as L1 and L2, are embedded in a two-dimensional square grid of physical qubits. The two patches are positioned such that their logical $X$ operators face each other and are connected by seven ancilla qubits (gray circles), enabling the implementation of lattice-merge (bottom to middle) and lattice-split (middle to top) operations. The logical operators for L1, L2 and the merged logical qubit L3 are represented as string operators acting on the data qubits enclosed within the respective boxes.
\textbf{c}, Cumulative distributions of physical operation errors. Blue: Pauli errors for single-qubit gates; gold: Pauli errors for CZ gates; black: Pauli errors for idling data qubits during measurement; green: readout assignment errors for two-state discrimination ($\ket{0}$, $\ket{1}$); red: readout assignment errors for three-state discrimination ($\ket{0}$, $\ket{1}$, $\ket{2}$). Dashed vertical lines indicate median values.
\textbf{d}, Logical memory performance of the surface codes. Two logical qubits are initialized in the state $\ket{0_\mathrm{L}}$, and then preserved for $t$ SECs. 
For each data point, we take $1 \times 10^6$ experimental repetitions and discard all runs of the experiment where a qubit is detected in a leaked state $\ket{2}$. The extracted logical error rates $\varepsilon_\mathrm{L1}$ and $\varepsilon_\mathrm{L2}$ are $0.0365(2)$ and $0.0282(1)$, respectively. 
}
\label{fig1}
\end{figure*}

Our quantum processor comprises a two-dimensional array of superconducting transmon qubits with tunable nearest-neighbor coupling (Fig.~\ref{fig1}a). Within this array, we define two logical qubits, L1 and L2, each encoded in a distance-three surface code. Each logical qubit is hosted on a patch consisting of nine data qubits ($\mathrm{D}j$) and eight syndrome qubits ($\mathrm{X}j$ and $\mathrm{Z}j$). These syndrome qubits measure X- and Z-type stabilizers on adjacent data qubits; for instance, qubit Z1 measures the stabilizer $S_{\mathrm{Z}1}=Z_{\mathrm{D}0}Z_{\mathrm{D}1}Z_{\mathrm{D}3}Z_{\mathrm{D}4}$. The logical space of each patch is defined by the simultaneous $+1$ eigenspace of all stabilizers, with the remaining degree of freedom encoding the logical information. We define the logical Pauli operators for L1 as $X_{\mathrm{L1}}=X_{\mathrm{D6}}X_{\mathrm{D7}}X_{\mathrm{D8}}$ and $Z_{\mathrm{L1}}=Z_{\mathrm{D0}}Z_{\mathrm{D3}}Z_{\mathrm{D6}}$; 
for L2, the operators are $X_{\mathrm{L2}}=X_{\mathrm{D12}}X_{\mathrm{D13}}X_{\mathrm{D14}}$ and $Z_{\mathrm{L2}}=Z_{\mathrm{D12}}Z_{\mathrm{D15}}Z_{\mathrm{D18}}$. To preserve the logical information, we perform repeated SECs, with each cycle comprising a full round of stabilizer measurements.

Within the lattice surgery framework, quantum computation is driven by measuring logical Pauli operators. To implement a fault-tolerant $X_{\mathrm{L1}}X_{\mathrm{L2}}$ measurement ($M_{XX}$), we position the two disjoint code patches such that their logical $X$ operators face each other. Between the two patches, we introduce three ancilla data qubits (D9, D10, D11) and four ancilla syndrome qubits (X4, X5, X6, X7), all initialized in the state $\ket{0}$. We then \textit{merge} L1 and L2 into a single surface code by dynamically reconfiguring the syndrome extraction circuits in the subsequent SECs. Specifically, the boundary Z-type stabilizer measured by Z3 (Z4) is extended from $Z_{\mathrm{D6}} Z_{\mathrm{D7}}$ ($Z_{\mathrm{D13}} Z_{\mathrm{D14}}$) to $Z_{\mathrm{D6}} Z_{\mathrm{D7}} Z_{\mathrm{D9}} Z_{\mathrm{D10}}$ ($Z_{\mathrm{D10}} Z_{\mathrm{D11}} Z_{\mathrm{D13}} Z_{\mathrm{D14}}$). Simultaneously, the four ancilla syndrome qubits measure newly established X-type stabilizers that act across the boundary data qubits of L1, L2, and the ancilla data qubits. This lattice-merge operation explicitly defines an elongated $7 \times 3$ rotated surface code, denoted as L3. Crucially, the operator $X_{\mathrm{L1}} X_{\mathrm{L2}}$ is equivalent to the product of the four newly established X-type stabilizers, $\prod_{j=4}^{7} S_{\mathrm{X}j}$; thus, the measurement outcome is extracted from the parity of these syndrome records, and the joint system is projected into an eigenspace of $X_{\mathrm{L1}} X_{\mathrm{L2}}$. Following the merge, the original two logical qubits can be recovered by performing a \textit{split} operation: we measure the three ancilla data qubits in the $Z$ basis, cease the X-type stabilizer measurements associated with the four ancilla syndrome qubits, and restore the boundary Z-type stabilizers corresponding to Z3 and Z4 from weight-four back to weight-two. As demonstrated in previous experiments~\cite{Besedin2026Nat.Phys.} and the following section, the lattice-split operation itself can be used to generate entangled states.

The physical performance of the device is summarized in Fig.~\ref{fig1}c, showing cumulative distributions of error probabilities for relevant operations. The median gate fidelities are $0.9995$ and $0.996$ for single-qubit gates and two-qubit controlled-$\pi$ phase (CZ) gates, respectively. The median Pauli error for data qubits idling during a measurement cycle is $0.024$. Readout assignment errors are $0.025$ for two-state discrimination ($\ket{0}$, $\ket{1}$) and $0.049$ for three-state discrimination ($\ket{0}$, $\ket{1}$, $\ket{2}$).
To benchmark logical memory performance, we conduct repeated SECs on the two disjoint code patches. By fitting the decay of logical fidelity~\cite{GoogleQuantumAI2023Nature} to an exponential function of the cycle number, we extract per-cycle logical error rates of $\varepsilon_\mathrm{L1}=0.0365(2)$ and $\varepsilon_\mathrm{L2}=0.0282(1)$ (Fig.~\ref{fig1}d). The difference between the two values reflects variation in physical-qubit performance across the device.

\section{Generation of a logical Bell state}

\begin{figure*}[!htbp]
\centering
\includegraphics[width=\textwidth]{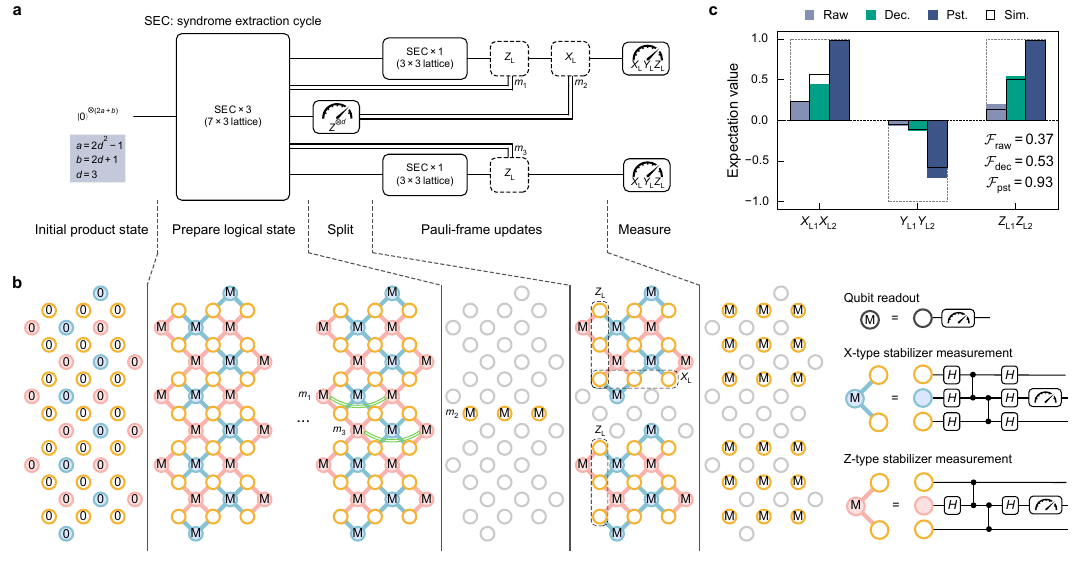}
\caption{
\textbf{Characterization of the logical Bell state generated via lattice split.} 
\textbf{a}, Simplified circuit for preparing a logical Bell state. The system is initialized by preparing all data qubits in the logical state $\ket{0_\mathrm{L3}}$ of a $7 \times 3$ merged surface code. The lattice is subsequently split into two entangled $3 \times 3$ code patches by measuring the ancilla data qubits. Pauli-frame updates, conditioned on the intermediate measurement outcomes ($m_1$, $m_2$, and $m_3$; see below), are applied during post-processing (dashed boxes).
\textbf{b}, Physical operations corresponding to each step in the circuit. Each SEC consists of four layers of CZ gates followed by a round of measurements on the syndrome qubits. To perform the lattice split, all ancilla data qubits are measured in the $Z$ basis, after which all ancilla qubits remain inactive. The intermediate outcomes $m_1$, $m_2$, and $m_3$ are obtained from the final round of syndrome extraction measurements of L3 and the subsequent measurement of the ancilla data qubits.
\textbf{c}, Measured expectation values of the logical Pauli observables for the prepared Bell state. The results are presented for raw data (purple), decoded data (green), and post-selected data conditioned on trivial error syndromes (blue). Dashed and solid wireframes show the ideal values and numerical simulation results, respectively.
}
\label{fig2}
\end{figure*}

A key practical application of lattice surgery is the deterministic generation of entanglement between logical qubits. We demonstrate this using a lattice-split protocol, as illustrated in Fig.~\ref{fig2}a-b. Initially, all data qubits are prepared in the state $\ket{0}$. We then execute three rounds of SECs, wherein syndrome qubits measure the stabilizers of the merged surface code, L3. These stabilizer measurements project the data qubits into the logical state $\ket{0_\mathrm{L3}}$ while providing the syndrome records necessary for fault-tolerant error decoding and correction. 
Following the third SEC, we split the merged code into L1 and L2 by measuring out the ancilla data qubits, thereafter reconfiguring the syndrome extraction circuits for two isolated code patches.
To make the split operation deterministic, we apply Pauli-frame updates conditioned on the intermediate measurement outcomes. 
Specifically, if the measurement of stabilizers during the third SEC yields an outcome of $-1$ for $S_{\mathrm{X}4}S_{\mathrm{X}5}$ ($S_{\mathrm{X}6}S_{\mathrm{X}7}$), we effectively apply a logical $Z$ gate to L1 (L2) by updating the Pauli frame; if the measurement of $Z_{\mathrm{D}9}$ yields $-1$, a logical $X$ gate is effectively applied to L1.
Accounting for these Pauli-frame updates, an X-type lattice split transforms the logical state according to~\cite{Horsman2012NewJ.Phys., Besedin2026Nat.Phys.}
\begin{equation}
\alpha\ket{+_\mathrm{L3}}+\beta\ket{-_\mathrm{L3}}
\;\to\;
\alpha\ket{+_\mathrm{L1}+_\mathrm{L2}}+\beta\ket{-_\mathrm{L1}-_\mathrm{L2}},
\label{eq:split_equation}
\end{equation}
where $\ket{\pm_\mathrm{L}}= \left(\ket{0_\mathrm{L}}\pm \ket{1_\mathrm{L}}\right) / {\sqrt{2}}$. 
Given our initialization in the state $\ket{0_\mathrm{L3}}$, which corresponds to $\alpha = \beta = 1 / {\sqrt{2}}$ in the logical $X$ basis, the resulting state is a logical Bell state, 
$\ket{\Phi^+_{\mathrm{L}}} = \left( \ket{0_{\mathrm{L}1} 0_{\mathrm{L}2}} + \ket{1_{\mathrm{L}1} 1_{\mathrm{L}2}} \right) / {\sqrt{2}}$.

To characterize the fidelity of the generated Bell state, we measure the expectation values of three two-qubit logical operators $X_{\mathrm{L}1}X_{\mathrm{L}2}$, $Y_{\mathrm{L}1}Y_{\mathrm{L}2}$, and $Z_{\mathrm{L}1}Z_{\mathrm{L}2}$ (Fig.~\ref{fig2}c). The state fidelity is calculated as~\cite{Erhard2021Nature}
\begin{equation}
\mathcal{F} = \frac{1}{4} \left( 1 + \left\langle X_{\mathrm{L}1}X_{\mathrm{L}2} \right\rangle - \left\langle Y_{\mathrm{L}1}Y_{\mathrm{L}2} \right\rangle + \left\langle Z_{\mathrm{L}1}Z_{\mathrm{L}2} \right\rangle \right).
\end{equation}
Before evaluating the expectation values, we discard runs in which leakage events are detected, retaining approximately $8.8 \times 10^5$ runs ($17.6\%$) out of $5 \times 10^6$ total experimental repetitions. The raw expectation values for the three logical operators are $0.224_{-2}^{+2}$, $-0.063_{-2}^{+2}$, and $0.200_{-2}^{+2}$, respectively, yielding a raw state fidelity of $\mathcal{F}_\mathrm{raw} = 0.372_{-1}^{+1}$. Here, the uncertainty represents a $95\%$ confidence interval calculated using bootstrapping.
Processing the logical outcomes together with syndrome records using a belief-matching decoder~\cite{Higgott2023Phys.Rev.X} significantly improves the expectation values of $X_{\mathrm{L1}}X_{\mathrm{L2}}$ and $Z_{\mathrm{L1}}Z_{\mathrm{L2}}$ to $0.449_{-2}^{+2}$ and $0.553_{-2}^{+2}$, respectively. 
By contrast, the decoded expectation value of $Y_{\mathrm{L1}}Y_{\mathrm{L2}}$ exhibits only a marginal improvement, from $-0.063_{-2}^{+2}$ to $-0.127_{-2}^{+2}$, as this specific measurement is non-fault-tolerant in the current implementation~\cite{Ye2023Phys.Rev.Lett.}, where only half of the X-type stabilizers and half of the Z-type stabilizers are calculated from the measurement outcomes of data qubits; the limited gain indicates significant measurement errors for $Y_{\mathrm{L1}}Y_{\mathrm{L2}}$.
These observations are consistent with numerical simulations using an error model extracted from benchmarking experiments. Consequently, the fidelity of the decoded state has a lower bound of $\mathcal{F}_{\mathrm{dec}}=0.532_{-1}^{+1}$, confirming the existence of genuine bipartite entanglement~\cite{Guhne2009Phys.Rep.}.

We further analyze the data using a detected-error-free post-selection protocol, retaining only those experimental runs where stabilizer measurements indicate no detected errors. The retained fraction is approximately $0.5\%$. For this post-selected data, the expectation values of the logical operators $X_{\mathrm{L1}}X_{\mathrm{L2}}$ and $Z_{\mathrm{L1}}Z_{\mathrm{L2}}$ reach $0.993_{-4}^{+4}$ and $0.997_{-3}^{+2}$, respectively, closely approaching their ideal value of $+1$. In contrast, the expectation value for $Y_{\mathrm{L1}}Y_{\mathrm{L2}}$ is $-0.712_{-20}^{+21}$, showing a notable departure from the ideal $-1$. From these expectation values, we establish a lower bound on the post-selected state fidelity of $\mathcal{F}_{\mathrm{pst}} = 0.925_{-5}^{+5}$.

\section{Deutsch-Jozsa algorithm on logical qubits}

\begin{figure}[tbp]
\centering
\includegraphics[width=3.5 in]{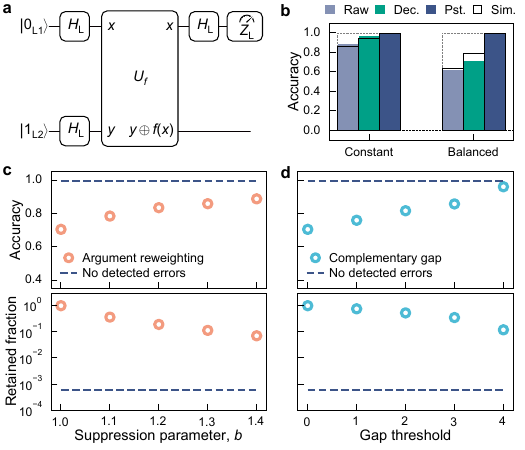}
\caption{
\textbf{Logical implementation of the Deutsch-Jozsa algorithm.}
\textbf{a}, Simplified circuit diagram of the Deutsch-Jozsa algorithm, which is proposed to distinguish between constant and balanced functions in a single trial. In this two-qubit implementation, the quantum oracle $U_f$ is efficiently realized via lattice surgery between two surface-code logical qubits. 
\textbf{b}, Measured accuracy of the logical algorithm for both constant and balanced functions. Dashed and solid wireframes represent the ideal values and simulation results, respectively.
\textbf{c}-\textbf{d}, Effect of soft post-selection based on argument reweighting (\textbf{c}) and complementary gap (\textbf{d}). Increasing the suppression parameter or gap threshold yields a monotonic improvement in logical accuracy at the cost of a reduced retained fraction. Blue dashed lines correspond to the detected-error-free post-selection approach, selecting runs with no detected errors throughout the process.
}
\label{fig3}
\end{figure}

To demonstrate the computational capabilities of lattice surgery beyond state preparation, we implement the two-qubit Deutsch-Jozsa algorithm using logical qubits. This algorithm distinguishes constant Boolean functions (which return the same value for all inputs) from balanced functions (which return $0$ for half the inputs and $1$ for the other half) in a single query. The logical circuit is shown in Fig.~\ref{fig3}a. Two logical qubits, L1 and L2, are initialized in the states $\ket{0_{\mathrm{L}1}}$ and $\ket{1_{\mathrm{L}2}}$, respectively, with logical Hadamard gates applied to prepare the superposition states $\ket{+_{\mathrm{L}1}}$ and $\ket{-_{\mathrm{L}2}}$. The implementation of the oracle $U_f$ depends on the function class: constant functions require no operation, while balanced functions are realized via lattice surgery; see Section~{S3} of Supplementary Information (SI). Finally, a logical Hadamard gate is applied to L1, followed by measurement. In the experiment, we simplify this procedure by directly preparing L1 and L2 in the $\ket{+_{\mathrm{L}1}}$ and $\ket{-_{\mathrm{L}2}}$ states and measuring L1 in the logical $X$ basis.

Ideally, the algorithm outputs a measurement outcome of $+1$ for constant functions and $-1$ for balanced functions. Fig.~\ref{fig3}b shows the measured classification accuracies for both scenarios. Error correction substantially improves the classification accuracy over the raw results: from $0.886_{-1}^{+1}$ to $0.970_{-1}^{+1}$ for the constant case, and from $0.623_{-1}^{+1}$ to $0.716_{-1}^{+1}$ for the balanced case. The higher accuracy observed in the constant case is attributed to the absence of lattice-surgery operations. Finally, as expected, applying detected-error-free post-selection yields near-unity accuracies of $0.99974_{-4}^{+4}$ and $0.995_{-7}^{+5}$ for the constant and balanced cases, respectively.

However, detected-error-free post-selection is highly inefficient, retaining only $0.06\%$ of experimental repetitions for the balanced function. To better explore the trade-off between classification accuracy and retained data fraction, we employ two more sophisticated post-selection strategies integrated with the minimum-weight perfect-matching (MWPM) decoder~\cite{Gidney2021Quantum}. In the argument-reweighting approach~\cite{Xie2026simple}, a second decoding round is performed using a modified error model. Specifically, the error probabilities assigned to edges constituting the optimal matching chains identified in the first round are suppressed ($p \rightarrow p^b$). A run is retained only if the two decoding rounds yield the same matching chains; otherwise, it is discarded. As the suppression parameter $b$ increases, classification accuracy improves at the cost of a lower retained fraction (Fig.~\ref{fig3}c). In the second approach, we utilize the complementary gap~\cite{ efficient_hutter_2014,faulttolerant_bombin_2024,Gidney2023yokedsurfacecodes,mitigating_smith_2024}, defined as the weight difference between the overall most likely error configuration and the most likely one belonging to a competing logical class. This method accepts only those runs with a complementary gap exceeding a specified threshold. Setting a larger threshold yields higher classification accuracy but fewer accepted runs (Fig.~\ref{fig3}d). Notably, this method reaches an accuracy of $0.960_{-1}^{+1}$ while retaining approximately $12\%$ of the data.

\section{Implementation and benchmarking of non-Clifford logical rotations}

\begin{figure*}[tbp]
\centering
\includegraphics[width=\textwidth]{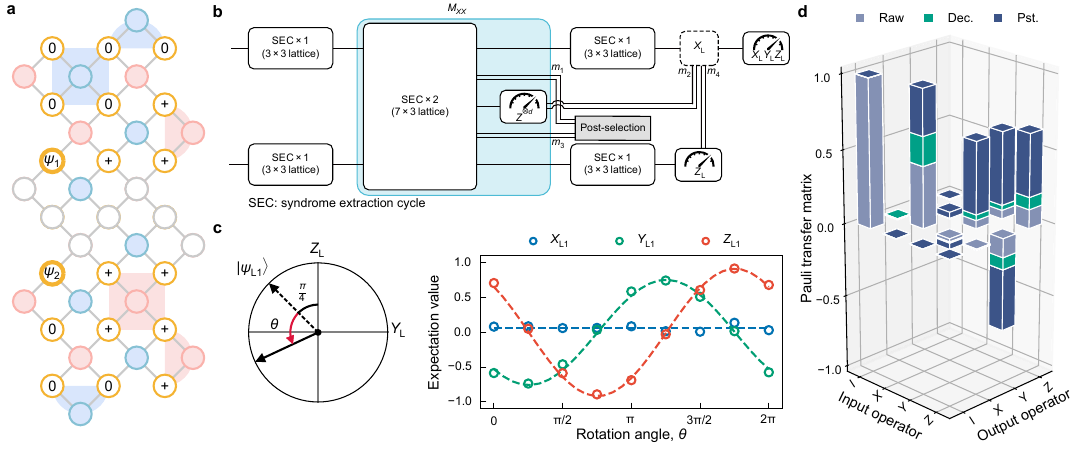}
\caption{
\textbf{Implementation of logical gate teleportation via lattice surgery.}
\textbf{a}, Protocol for logical magic state preparation. For each surface-code patch, the unique physical data qubit participating in both $X_\mathrm{L}$ and $Z_\mathrm{L}$ is initialized in the target physical magic state ($\ket{\psi_{1}}$ or $\ket{\psi_{2}}$). The remaining data qubits are prepared in $\ket{0}$ or $\ket{+}$, such that a subsequent round of syndrome extraction projects the code into the desired logical magic state ($\ket{\psi_\mathrm{L1}}$ or $\ket{\psi_\mathrm{L2}}$).
\textbf{b}, Circuit diagram for an arbitrary rotation around the logical $X$ axis. Lattice merge and split operations between two surface-code patches implement a joint parity measurement $M_{XX}$. Because the measurement outcome $m=m_1m_3$ indicates the rotation direction, we post-select the experimental runs where $m=1$ for the desired rotation gate.
The magic-state logical qubit L2 is then measured, with its outcome processed to apply a conditional Pauli-frame update on L1 (dashed box).
\textbf{c}, Measured expectation values of the logical Pauli operators as functions of the rotation angle $\theta$. The logical data qubit L1 is initialized in an $H$-type magic state lying in the $Y_\mathrm{L}Z_\mathrm{L}$ plane of the Bloch sphere (dashed arrow). After the lattice surgery and post-processing procedures, the final state (solid arrow) is characterized by measuring the expectation values of the logical Pauli operators $X_\mathrm{L}$ (blue), $Y_\mathrm{L}$ (green), and $Z_\mathrm{L}$ (red). Dashed lines indicate fits to the experimental data.
\textbf{d}, Pauli transfer matrix (PTM) of the logical $R_{X}(\pi/4)$ gate. The experimental PTM is reconstructed using raw (purple), decoded (green), and post-selected (blue) logical outcomes. 
}
\label{fig4}
\end{figure*}

Universal quantum computation requires the ability to perform arbitrary rotations, necessitating the inclusion of non-Clifford elements in the logical gate set. We can realize such operations by consuming high-fidelity magic states through a gate-teleportation protocol~\cite{Bravyi2005Phys.Rev.A}. Here, we demonstrate non-Clifford logical gates by utilizing L1 as the target logical qubit and preparing the required magic state on L2.

The two logical qubits are initialized using the protocol~\cite{Ye2023Phys.Rev.Lett.} shown in Fig.~\ref{fig4}a. We prepare L1 in the state $\ket{\psi_{\mathrm{L}1}} = \alpha \ket{+_{\mathrm{L}1}} + \beta \ket{-_{\mathrm{L}1}}$ as the input to the logical gate, while L2 is prepared in the magic state $\ket{\psi_{\mathrm{L}2}} = (\ket{+_{\mathrm{L}2}} + e^{i\theta}\ket{-_{\mathrm{L}2}})/\sqrt{2}$ required for the logical $R_{X}(\theta)$ rotation. State preparation proceeds by initializing the data qubit D6 (D12), located at the intersection of the logical $X$ and $Z$ operators, in the physical state $\ket{\psi_1} = \alpha \ket{+} + \beta \ket{-}$ ($\ket{\psi_2} = \left(\ket{+} + e^{i\theta}\ket{-}\right)/\sqrt{2}$). The other data qubits participating in the logical $X$ ($Z$) operators are initialized to $\ket{+}$ ($\ket{0}$); subject to these constraints, the remaining data qubits are set to either $\ket{+}$ or $\ket{0}$ to maximize the number of stabilizers yielding a deterministic $+1$ measurement outcome in the absence of errors. While this initialization is non-fault-tolerant, the resulting magic states can be further distilled to achieve the high fidelities required for large-scale computation~\cite{Bravyi2005Phys.Rev.A, magic_litinski_2019, Fowler2012Phys.Rev.A}.

In this experiment, we directly employ raw magic states to realize arbitrary logical $R_X(\theta)$ rotations on L1 (Fig.~\ref{fig4}b). Gate teleportation is implemented via a joint $M_{XX}$ measurement between L1 and L2, realized through lattice-merge and lattice-split operations. It is worth noting that, for the merged code, we apply two rounds of SECs rather than three in order to keep sufficient experimental runs after detected-error-free post-selection on syndromes; consequently, this implementation is not fully fault-tolerant. Following this joint measurement, we measure L2 and apply a conditional logical $X$ gate on L1. The $M_{XX}$ measurement outcome determines the rotation direction: an outcome of $+1$ induces the target rotation by $\theta$, while an outcome of $-1$ results in a rotation by $-\theta$. Although the rotation direction can, in principle, be corrected using active feedback, we here adopt a post-selection strategy, accepting only those experimental runs where the joint measurement yields a $+1$ outcome.

Fig.~\ref{fig4}c illustrates the successful execution of the non-Clifford logical rotation. We initialize L1 in the state $\ket{\psi_{\mathrm{L}1}} = (\ket{+_{\mathrm{L}1}} + e^{i\pi / 4} \ket{-_{\mathrm{L}1}})/\sqrt{2}$, apply the logical $R_{X}(\theta)$ gate, and measure the expectation values of logical Pauli operators as functions of the rotation angle $\theta$. 
After applying detected-error-free post-selection, we fit the expectation values to sinusoidal functions. The results reveal a reduced oscillation amplitude for logical $Y$ ($0.757$) compared to logical $Z$ ($0.916$). This asymmetry arises because our current implementation of the logical $Y$ measurement is non-fault-tolerant, leading to higher effective readout noise for that observable.

To rigorously characterize a non-Clifford operation, we perform quantum process tomography on the logical $R_{X}(\pi/4)$ gate. Fig.~\ref{fig4}d displays the reconstructed Pauli transfer matrix (PTM)~\cite{Merkel2013Phys.Rev.A, Marques2022Nat.Phys.} for this logical gate. Although performance is currently limited by infidelities in both the initial logical-state preparation and the lattice-surgery operations, the error-corrected results show a modest improvement over the raw data. Notably, detected-error-free post-selection achieves an average logical-gate fidelity~\cite{Marques2022Nat.Phys.} of $0.943_{-9}^{+10}$, demonstrating that a protocol based on lattice surgery can realize high-fidelity universal gates.

\section{Discussion and outlook}

In summary, we have demonstrated lattice-surgery operations between two logical qubits protected by the distance-three rotated surface code on a superconducting processor. Specifically, we have realized logical Bell states with error-corrected fidelities surpassing the entanglement threshold, executed the Deutsch-Jozsa algorithm at the logical level, and attained arbitrary-angle rotations via magic-state injection and gate teleportation. Together, these results provide a comprehensive validation of lattice surgery and establish the essential building blocks required for universal fault-tolerant quantum computation in planar architectures.

Our work motivates several immediate directions for future research. 
On the hardware side, improvements in physical-gate fidelity, leakage suppression, qubit-reset capability~\cite{Magnard2018Phys.Rev.Lett., McEwen2021Nat.Commun., Marques2023Phys.Rev.Lett.}, and readout discrimination should directly enhance the performance of logical operations while reducing reliance on post-selection. 
On the coding side, scaling to larger code distances will enable direct tests of whether the fidelity of lattice-surgery operations improves with code distance, which is a defining hallmark of fault tolerance. Equally important is the integration of real-time decoding~\cite{PRXQuantum.4.040344,skoric2023parallel,10.1063/1.1499754,Caune2024arxiv} and feedback operations~\cite{Baumer2024PRXQuantum}, which would enable active error correction.
Beyond the specific demonstrations reported here, the same framework can be extended to logical state teleportation~\cite{Zou2026teleportationtransitionsurfacecodes}, logical randomized benchmarking~\cite{Lacroix2025Nature}, and magic state distillation~\cite{SalesRodriguez2025Nature}. 
We therefore anticipate that lattice surgery will play a central role in the evolution from isolated quantum memories to programmable fault-tolerant superconducting quantum processors.

\vspace{10pt}

\noindent {\bf Acknowledgments} The device was fabricated at the Micro-Nano Fabrication Center of Zhejiang University. 
We acknowledge the support from the National Natural Science Foundation of China (grant nos.~92565301, 12404574, 12404570, and 12225507), the Zhejiang Leading Goose (Lingyan) Project (grant no.~2026C02A2004), and the Zhejiang Pioneer (Jianbing) Project (grant no.~2025C01019). 
H. Wang is also supported by the New Cornerstone Science Foundation through the XPLORER PRIZE.

\vspace{10pt}

\noindent {\bf Author contributions} Y.L., H.X., and P.Z. conceived the project; Y.~Wang and F.S. carried out the experiments and analyzed the experimental data under the supervision of P.Z. and H.~Wang; H.L. fabricated the device supervised by H.~Wang; Y.L., P.Z., Y.~Wang, H.X., F.S., and H.~Wang co-wrote the manuscript; H.~Wang, P.Z., Z.W., C.S., Q.G., Y.~Wang, F.S., A.Z., Y.G., C.Z., X.~Zhu, F.J., Y.Z., N.W., Z.C., Z.B., Z.Z., J.Z., G.L., J.-N.Y., Y.~Han, Y.~He, J.S., Han~Wang, J.H., X.~Zhang, S.Z., H.D., J.D., Y.~Wu, and Z.S. contributed to the experimental setup. All authors contributed to the discussions of the results.

\bibliography{references}

@article{Kitaev2003Ann.Phys.,
  title = {Fault-tolerant quantum computation by anyons},
  author = {A.Yu. Kitaev},
  journal = {Ann. Phys.},
  volume = {303},
  number = {1},
  pages = {2-30},
  year = {2003},
  issn = {0003-4916},
  doi = {https://doi.org/10.1016/S0003-4916(02)00018-0},
  url = {https://www.sciencedirect.com/science/article/pii/S0003491602000180},
}

@article{Fowler2012Phys.Rev.A,
  title = {Surface Codes: Towards Practical Large-Scale Quantum Computation},
  shorttitle = {Surface Codes},
  author = {Fowler, Austin G. and Mariantoni, Matteo and Martinis, John M. and Cleland, Andrew N.},
  year = 2012,
  journal = {Phys. Rev. A},
  volume = {86},
  number = {3},
  pages = {032324},
  issn = {1050-2947, 1094-1622},
  doi = {10.1103/PhysRevA.86.032324},
  url = {https://link.aps.org/doi/10.1103/PhysRevA.86.032324}
}

@article{Krinner2022Nature,
  title = {Realizing Repeated Quantum Error Correction in a Distance-Three Surface Code},
  author = {Krinner, Sebastian and others},
  year = {2022},
  journal = {Nature},
  volume = {605},
  number = {7911},
  pages = {669--674},
  publisher = {Nature Publishing Group},
  issn = {1476-4687},
  doi = {10.1038/s41586-022-04566-8},
  url = {https://www.nature.com/articles/s41586-022-04566-8}
}

@article{He2025Phys.Rev.Lett.,
  title = {Experimental Quantum Error Correction below the Surface Code Threshold via All-Microwave Leakage Suppression},
  author = {He, Tan and others},
  year = {2025},
  journal = {Phys. Rev. Lett.},
  volume = {135},
  number = {26},
  pages = {260601},
  publisher = {American Physical Society},
  doi = {10.1103/rqkg-dw31},
  url = {https://link.aps.org/doi/10.1103/rqkg-dw31}
}

@article{Acharya2025Nature,
  title = {Quantum Error Correction below the Surface Code Threshold},
  author = {Acharya, Rajeev and others},
  year = {2025},
  journal = {Nature},
  volume = {638},
  number = {8052},
  pages = {920--926},
  publisher = {Nature Publishing Group},
  issn = {1476-4687},
  doi = {10.1038/s41586-024-08449-y},
}

@article{Ryan-Anderson2024Science,
  title = {High-Fidelity Teleportation of a Logical Qubit Using Transversal Gates and Lattice Surgery},
  author = {{Ryan-Anderson}, C. and others},
  year = {2024},
  journal = {Science},
  volume = {385},
  number = {6715},
  pages = {1327--1331},
  publisher = {American Association for the Advancement of Science},
  doi = {10.1126/science.adp6016},
  url = {https://www.science.org/doi/10.1126/science.adp6016}
}

@article{Bluvstein2024Nature,
  title = {Logical Quantum Processor Based on Reconfigurable Atom Arrays},
  author = {Bluvstein, Dolev and others},
  year = {2024},
  journal = {Nature},
  volume = {626},
  number = {7997},
  pages = {58--65},
  publisher = {Nature Publishing Group},
  issn = {1476-4687},
  doi = {10.1038/s41586-023-06927-3},
  url = {https://www.nature.com/articles/s41586-023-06927-3}
}

@article{Bluvstein2026Nature,
  title = {A Fault-Tolerant Neutral-Atom Architecture for Universal Quantum Computation},
  author = {Bluvstein, Dolev and others},
  year = {2026},
  journal = {Nature},
  volume = {649},
  number = {8095},
  pages = {39--46},
  publisher = {Nature Publishing Group},
  issn = {1476-4687},
  doi = {10.1038/s41586-025-09848-5},
  url = {https://www.nature.com/articles/s41586-025-09848-5}
}

@article{Horsman2012NewJ.Phys.,
  title = {Surface Code Quantum Computing by Lattice Surgery},
  author = {Horsman, Dominic and Fowler, Austin G and Devitt, Simon and Meter, Rodney Van},
  year = {2012},
  journal = {New J. Phys.},
  volume = {14},
  number = {12},
  pages = {123011},
  publisher = {IOP Publishing},
  issn = {1367-2630},
  doi = {10.1088/1367-2630/14/12/123011},
  url = {https://dx.doi.org/10.1088/1367-2630/14/12/123011}
}

@article{Lacroix2025Nature,
  title = {Scaling and Logic in the Colour Code on a Superconducting Quantum Processor},
  author = {Lacroix, N. and others},
  year = {2025},
  journal = {Nature},
  volume = {645},
  number = {8081},
  pages = {614--619},
  publisher = {Nature Publishing Group},
  issn = {1476-4687},
  doi = {10.1038/s41586-025-09061-4},
  url = {https://www.nature.com/articles/s41586-025-09061-4}
}

@article{Besedin2026Nat.Phys.,
  title = {Lattice Surgery Realized on Two Distance-Three Repetition Codes with Superconducting Qubits},
  author = {Besedin, Ilya and others},
  year = {2026},
  journal = {Nat. Phys.},
  volume = {22},
  number = {2},
  pages = {189--194},
  publisher = {Nature Publishing Group},
  issn = {1745-2481},
  doi = {10.1038/s41567-025-03090-6},
  url = {https://www.nature.com/articles/s41567-025-03090-6}
}

@misc{Xie2026simple,
  title = {Simple, Efficient, and Generic Post-Selection Decoding for qLDPC codes}, 
  author = {Haipeng Xie and Nobuyuki Yoshioka and Kento Tsubouchi and Ying Li},
  year = {2026},
  eprint = {2601.17757},
  archivePrefix = {arXiv},
  primaryClass = {quant-ph},
}

@misc{Gidney2023yokedsurfacecodes,
  title={Yoked surface codes}, 
  author={Craig Gidney and Michael Newman and Peter Brooks and Cody Jones},
  year={2023},
  eprint={2312.04522},
  archivePrefix={arXiv},
  primaryClass={quant-ph},
}

@article{Bravyi2005Phys.Rev.A,
  title = {Universal quantum computation with ideal Clifford gates and noisy ancillas},
  author = {Bravyi, Sergey and Kitaev, Alexei},
  journal = {Phys. Rev. A},
  volume = {71},
  issue = {2},
  pages = {022316},
  numpages = {14},
  year = {2005},
  month = {Feb},
  publisher = {American Physical Society},
  doi = {10.1103/PhysRevA.71.022316},
  url = {https://link.aps.org/doi/10.1103/PhysRevA.71.022316}
}

@article{Higgott2023Phys.Rev.X,
  title = {Improved Decoding of Circuit Noise and Fragile Boundaries of Tailored Surface Codes},
  author = {Higgott, Oscar and Bohdanowicz, Thomas C. and Kubica, Aleksander and Flammia, Steven T. and Campbell, Earl T.},
  year = {2023},
  journal = {Phys. Rev. X},
  volume = {13},
  number = {3},
  pages = {031007},
  publisher = {American Physical Society},
  doi = {10.1103/PhysRevX.13.031007},
  url = {https://link.aps.org/doi/10.1103/PhysRevX.13.031007}
}

@misc{Zou2026teleportationtransitionsurfacecodes,
  title = {Teleportation transition of surface codes on a superconducting quantum processor}, 
  author = {Yiren Zou and others},
  year = {2026},
  eprint = {2602.21293},
  archivePrefix = {arXiv},
  primaryClass = {quant-ph},
}

@article{GoogleQuantumAI2023Nature,
  title = {Suppressing Quantum Errors by Scaling a Surface Code Logical Qubit},
  author = {Acharya, Rajeev and others},
  year = {2023},
  journal = {Nature},
  volume = {614},
  number = {7949},
  pages = {676--681},
  publisher = {Nature Publishing Group},
  issn = {1476-4687},
  doi = {10.1038/s41586-022-05434-1},
  url = {https://www.nature.com/articles/s41586-022-05434-1}
}

@article{Zhang2025Nat.Commun.,
  title = {Demonstrating Quantum Error Mitigation on Logical Qubits},
  author = {Zhang, Aosai and others},
  year = {2026},
  journal = {Nat. Commun.},
  volume = {17},
  number = {1},
  pages = {1021},
  publisher = {Nature Publishing Group},
  issn = {2041-1723},
  doi = {10.1038/s41467-025-67768-4},
  url = {https://www.nature.com/articles/s41467-025-67768-4}
}

@article{Terhal2015Rev.Mod.Phys.,
  title = {Quantum error correction for quantum memories},
  author = {Terhal, Barbara M.},
  journal = {Rev. Mod. Phys.},
  volume = {87},
  issue = {2},
  pages = {307--346},
  numpages = {40},
  year = {2015},
  month = {Apr},
  publisher = {American Physical Society},
  doi = {10.1103/RevModPhys.87.307},
  url = {https://link.aps.org/doi/10.1103/RevModPhys.87.307}
}

@article{Litinski2019Quantum,
  title = {A Game of Surface Codes: Large-Scale Quantum Computing with Lattice Surgery},
  shorttitle = {A Game of Surface Codes},
  author = {Litinski, Daniel},
  year = {2019},
  journal = {Quantum},
  volume = {3},
  pages = {128},
  issn = {2521-327X},
  doi = {10.22331/q-2019-03-05-128},
  url = {http://arxiv.org/abs/1808.02892},
}

@misc{Bravyi1998quantum,
  title={Quantum codes on a lattice with boundary}, 
  author={S. B. Bravyi and A. Yu. Kitaev},
  year={1998},
  eprint={quant-ph/9811052},
  archivePrefix={arXiv},
}

@article{SalesRodriguez2025Nature,
  title = {Experimental Demonstration of Logical Magic State Distillation},
  author = {Sales Rodriguez, Pedro and others},
  year = {2025},
  journal = {Nature},
  volume = {645},
  number = {8081},
  pages = {620--625},
  publisher = {Nature Publishing Group},
  issn = {1476-4687},
  doi = {10.1038/s41586-025-09367-3},
  url = {https://www.nature.com/articles/s41586-025-09367-3}
}

@article{Shor1995Phys.Rev.A.,
  title = {Scheme for reducing decoherence in quantum computer memory},
  author = {Shor, Peter W.},
  journal = {Phys. Rev. A},
  volume = {52},
  issue = {4},
  pages = {R2493--R2496},
  numpages = {0},
  year = {1995},
  month = {Oct},
  publisher = {American Physical Society},
  doi = {10.1103/PhysRevA.52.R2493},
  url = {https://link.aps.org/doi/10.1103/PhysRevA.52.R2493}
}

@article{Calderbank1996Phys.Rev.A,
  title = {Good Quantum Error-Correcting Codes Exist},
  author = {Calderbank, A. R. and Shor, Peter W.},
  year = {1996},
  journal = {Phys. Rev. A},
  volume = {54},
  number = {2},
  pages = {1098--1105},
  publisher = {American Physical Society},
  doi = {10.1103/PhysRevA.54.1098},
  url = {https://link.aps.org/doi/10.1103/PhysRevA.54.1098}
}

@article{Steane1996Phys.Rev.Lett.,
  title = {Error Correcting Codes in Quantum Theory},
  author = {Steane, A. M.},
  year = {1996},
  journal = {Phys. Rev. Lett.},
  volume = {77},
  number = {5},
  pages = {793--797},
  publisher = {American Physical Society},
  doi = {10.1103/PhysRevLett.77.793},
  url = {https://link.aps.org/doi/10.1103/PhysRevLett.77.793}
}

@article{Knill1997Phys.Rev.A,
  title = {Theory of Quantum Error-Correcting Codes},
  author = {Knill, Emanuel and Laflamme, Raymond},
  year = {1997},
  journal = {Phys. Rev. A},
  volume = {55},
  number = {2},
  pages = {900--911},
  publisher = {American Physical Society},
  doi = {10.1103/PhysRevA.55.900},
  url = {https://link.aps.org/doi/10.1103/PhysRevA.55.900}
}

@article{Dennis2002J.Math.Phys.,
  title = {Topological Quantum Memory},
  author = {Dennis, Eric and Kitaev, Alexei and Landahl, Andrew and Preskill, John},
  year = {2002},
  journal = {J. Math. Phys.},
  volume = {43},
  number = {9},
  pages = {4452--4505},
  issn = {0022-2488},
  doi = {10.1063/1.1499754},
  url = {https://doi.org/10.1063/1.1499754}
}

@article{Erhard2021Nature,
  title = {Entangling Logical Qubits with Lattice Surgery},
  author = {Erhard, Alexander and others},
  year = {2021},
  journal = {Nature},
  volume = {589},
  number = {7841},
  pages = {220--224},
  publisher = {Nature Publishing Group},
  issn = {1476-4687},
  doi = {10.1038/s41586-020-03079-6},
  url = {https://www.nature.com/articles/s41586-020-03079-6}
}

@misc{Shor1997faulttolerantquantumcomputation,
  title = {Fault-tolerant quantum computation}, 
  author = {Peter W. Shor},
  year = {1997},
  eprint = {quant-ph/9605011},
  archivePrefix = {arXiv},
  primaryClass = {quant-ph},
}

@article{Magnard2018Phys.Rev.Lett.,
  title = {Fast and Unconditional All-Microwave Reset of a Superconducting Qubit},
  author = {Magnard, P. and others},
  year = {2018},
  journal = {Phys. Rev. Lett.},
  volume = {121},
  number = {6},
  pages = {060502},
  publisher = {American Physical Society},
  doi = {10.1103/PhysRevLett.121.060502},
  url = {https://link.aps.org/doi/10.1103/PhysRevLett.121.060502}
}

@article{McEwen2021Nat.Commun.,
  title = {Removing Leakage-Induced Correlated Errors in Superconducting Quantum Error Correction},
  author = {McEwen, M. and others},
  year = {2021},
  journal = {Nat. Commun.},
  volume = {12},
  number = {1},
  pages = {1761},
  issn = {2041-1723},
  doi = {10.1038/s41467-021-21982-y},
  url = {http://www.nature.com/articles/s41467-021-21982-y}
}

@article{Marques2023Phys.Rev.Lett.,
  title = {All-Microwave Leakage Reduction Units for Quantum Error Correction with Superconducting Transmon Qubits},
  author = {Marques, J. F. and others},
  year = {2023},
  journal = {Phys. Rev. Lett.},
  volume = {130},
  number = {25},
  pages = {250602},
  issn = {0031-9007, 1079-7114},
  doi = {10.1103/PhysRevLett.130.250602},
  url = {https://link.aps.org/doi/10.1103/PhysRevLett.130.250602}
}

@article{Baumer2024PRXQuantum,
  title = {Efficient Long-Range Entanglement Using Dynamic Circuits},
  author = {B{\"a}umer, Elisa},
  year = {2024},
  journal = {PRX Quantum},
  volume = {5},
  number = {3},
  pages = {030339},
  issn = {2691-3399},
  doi = {10.1103/PRXQuantum.5.030339},
  url = {https://link.aps.org/doi/10.1103/PRXQuantum.5.030339}
}

@article{Gidney2021Quantum,
  title = {Stim: a fast stabilizer circuit simulator},
  author = {Gidney, Craig},
  journal = {{Quantum}},
  issn = {2521-327X},
  volume = {5},
  pages = {497},
  month = {jul},
  year = {2021},
  doi = {10.22331/q-2021-07-06-497},
  url = {http://arxiv.org/abs/2103.02202}
}

@article{Gulde2003Nature,
  title = {Implementation of the Deutsch-Jozsa Algorithm on an Ion-Trap Quantum Computer},
  author = {Gulde, Stephan and others},
  year = {2003},
  journal = {Nature},
  volume = {421},
  number = {6918},
  pages = {48--50},
  issn = {1476-4687},
  doi = {10.1038/nature01336},
  url = {https://doi.org/10.1038/nature01336}
}

@article{Shen2026Nat.Phys.,
  title = {A Bucket-Brigade Quantum Random Access Memory},
  author = {Shen, Fanhao and others},
  year = {2026},
  journal = {Nat. Phys.},
  volume = {22},
  pages = {745--750},
  publisher = {Nature Publishing Group},
  issn = {1745-2481},
  doi = {10.1038/s41567-026-03218-2},
  url = {https://www.nature.com/articles/s41567-026-03218-2}
}

@article{Motzoi2009Phys.Rev.Lett.,
  title = {Simple Pulses for Elimination of Leakage in Weakly Nonlinear Qubits},
  author = {Motzoi, F. and Gambetta, J. M. and Rebentrost, P. and Wilhelm, F. K.},
  year = {2009},
  journal = {Phys. Rev. Lett.},
  volume = {103},
  number = {11},
  pages = {110501},
  issn = {0031-9007, 1079-7114},
  doi = {10.1103/PhysRevLett.103.110501},
  url = {https://link.aps.org/doi/10.1103/PhysRevLett.103.110501}
}

@article{Wang2026Nat.Phys.,
  title={Demonstration of low-overhead quantum error correction codes},
  author={Wang, Ke and others},
  journal={Nat. Phys.},
  year={2026},
  month={Feb},
  day={01},
  volume={22},
  number={2},
  pages={308-314},
  issn={1745-2481},
  doi={10.1038/s41567-025-03157-4},
  url={https://doi.org/10.1038/s41567-025-03157-4}
}

@article{Ye2023Phys.Rev.Lett.,
  title = {Logical Magic State Preparation with Fidelity beyond the Distillation Threshold on a Superconducting Quantum Processor},
  author = {Ye, Yangsen and others},
  year = {2023},
  journal = {Phys. Rev. Lett.},
  volume = {131},
  number = {21},
  pages = {210603},
  publisher = {American Physical Society},
  doi = {10.1103/PhysRevLett.131.210603},
  url = {https://link.aps.org/doi/10.1103/PhysRevLett.131.210603}
}

@article{Breuckmann2021PRXQuantum,
  title = {Quantum Low-Density Parity-Check Codes},
  author = {Breuckmann, Nikolas P. and Eberhardt, Jens Niklas},
  journal = {PRX Quantum},
  volume = {2},
  issue = {4},
  pages = {040101},
  numpages = {19},
  year = {2021},
  month = {Oct},
  publisher = {American Physical Society},
  doi = {10.1103/PRXQuantum.2.040101},
  url = {https://link.aps.org/doi/10.1103/PRXQuantum.2.040101}
}

@article{Bravyi2024Nature,
  title = {High-Threshold and Low-Overhead Fault-Tolerant Quantum Memory},
  author = {Bravyi, Sergey and others},
  year = {2024},
  journal = {Nature},
  volume = {627},
  number = {8005},
  pages = {778--782},
  publisher = {Nature Publishing Group},
  issn = {1476-4687},
  doi = {10.1038/s41586-024-07107-7},
  url = {https://www.nature.com/articles/s41586-024-07107-7}
}

@article{Ni2023Nature,
  title = {Beating the Break-Even Point with a Discrete-Variable-Encoded Logical Qubit},
  author = {Ni, Zhongchu and others},
  year = {2023},
  journal = {Nature},
  volume = {616},
  number = {7955},
  pages = {56--60},
  publisher = {Nature Publishing Group},
  issn = {1476-4687},
  doi = {10.1038/s41586-023-05784-4},
  url = {https://www.nature.com/articles/s41586-023-05784-4}
}

@article{Marques2022Nat.Phys.,
  title = {Logical-Qubit Operations in an Error-Detecting Surface Code},
  author = {Marques, J. F. and others},
  year = {2022},
  journal = {Nat. Phys.},
  volume = {18},
  number = {1},
  pages = {80--86},
  issn = {1745-2473, 1745-2481},
  doi = {10.1038/s41567-021-01423-9},
  url = {https://www.nature.com/articles/s41567-021-01423-9}
}

@article{Merkel2013Phys.Rev.A,
  title = {Self-consistent quantum process tomography},
  author = {Merkel, Seth T. and Gambetta, Jay M. and Smolin, John A. and Poletto, Stefano and C\'orcoles, Antonio D. and Johnson, Blake R. and Ryan, Colm A. and Steffen, Matthias},
  journal = {Phys. Rev. A},
  volume = {87},
  issue = {6},
  pages = {062119},
  numpages = {9},
  year = {2013},
  month = {Jun},
  publisher = {American Physical Society},
  doi = {10.1103/PhysRevA.87.062119},
  url = {https://link.aps.org/doi/10.1103/PhysRevA.87.062119}
}

@article{Guhne2009Phys.Rep.,
  title = {Entanglement detection},
  author = {Otfried Gühne and Géza Tóth},
  journal = {Phys. Rep.},
  volume = {474},
  number = {1},
  pages = {1-75},
  year = {2009},
  issn = {0370-1573},
  doi = {https://doi.org/10.1016/j.physrep.2009.02.004},
  url = {https://www.sciencedirect.com/science/article/pii/S0370157309000623},
}

@misc{Caune2024arxiv,
  title={Demonstrating real-time and low-latency quantum error correction with superconducting qubits}, 
  author={Laura Caune and others},
  year={2024},
  eprint={2410.05202},
  archivePrefix={arXiv},
  primaryClass={quant-ph},
}

@article{Krantz2019Appl.Phys.Rev.,
  title = {A Quantum Engineer's Guide to Superconducting Qubits},
  author = {Krantz, P. and Kjaergaard, M. and Yan, F. and Orlando, T. P. and Gustavsson, S. and Oliver, W. D.},
  year = {2019},
  journal = {Appl. Phys. Rev.},
  volume = {6},
  number = {2},
  pages = {021318},
  issn = {1931-9401},
  doi = {10.1063/1.5089550},
  url = {http://aip.scitation.org/doi/10.1063/1.5089550}
}

@article{Kjaergaard2020Annu.Rev.Condens.MatterPhys.,
  title = {Superconducting Qubits: Current State of Play},
  shorttitle = {Superconducting {{Qubits}}},
  author = {Kjaergaard, Morten and others},
  year = {2020},
  journal = {Annu. Rev. Condens. Matter Phys.},
  volume = {11},
  number = {1},
  pages = {369-395},
  doi = {10.1146/annurev-conmatphys-031119-050605},
}

@article{PRXQuantum.4.040344,
  title = {Scalable Surface-Code Decoders with Parallelization in Time},
  author = {Tan, Xinyu and Zhang, Fang and Chao, Rui and Shi, Yaoyun and Chen, Jianxin},
  journal = {PRX Quantum},
  volume = {4},
  issue = {4},
  pages = {040344},
  numpages = {16},
  year = {2023},
  month = {Dec},
  publisher = {American Physical Society},
  doi = {10.1103/PRXQuantum.4.040344},
  url = {https://link.aps.org/doi/10.1103/PRXQuantum.4.040344}
}

@article{skoric2023parallel,
  author={Skoric, Luka and Browne, Dan E. and Barnes, Kenton M. and Gillespie, Neil I. and Campbell, Earl T.},
  title={Parallel window decoding enables scalable fault tolerant quantum computation},
  journal={Nat. Commun.},
  year={2023},
  month={Nov},
  day={03},
  volume={14},
  number={1},
  pages={7040},
  issn={2041-1723},
  doi={10.1038/s41467-023-42482-1},
  url={https://doi.org/10.1038/s41467-023-42482-1}
}

@article{10.1063/1.1499754,
  author = {Dennis, Eric and Kitaev, Alexei and Landahl, Andrew and Preskill, John},
  title = {Topological quantum memory},
  journal = {J. Math. Phys.},
  volume = {43},
  number = {9},
  pages = {4452-4505},
  year = {2002},
  month = {09},
  issn = {0022-2488},
  doi = {10.1063/1.1499754},
  url = {https://doi.org/10.1063/1.1499754},
}

@article{10.1063/1.1716296,
    author = {Meiboom, S. and Gill, D.},
    title = {Modified spin-echo method for measuring nuclear relaxation times},
    journal = {Review of Scientific Instruments},
    volume = {29},
    number = {8},
    pages = {688-691},
    year = {1958},
    month = {08},
    issn = {0034-6748},
    doi = {10.1063/1.1716296},
    url = {https://doi.org/10.1063/1.1716296},
}

@article{arute2019quantum,
  title={Quantum supremacy using a programmable superconducting processor},
  author={Arute, Frank and others},
  journal={Nature},
  volume={574},
  number={7779},
  pages={505--510},
  year={2019},
  month = {10},
  issn = {1476-4687},
  doi = {10.1038/s41586-019-1666-5},
  url = {https://doi.org/10.1038/s41586-019-1666-5},
}

@article{efficient_hutter_2014,
  title = {Efficient {{Markov}} Chain {{Monte Carlo}} Algorithm for the Surface Code},
  author = {Hutter, Adrian and Wootton, James R. and Loss, Daniel},
  year = 2014,
  journal = {Phys. Rev. A},
  volume = {89},
  number = {2},
  pages = {022326},
  doi = {10.1103/PhysRevA.89.022326}
}

@article{faulttolerant_bombin_2024,
  title = {Fault-{{Tolerant Postselection}} for {{Low-Overhead Magic State Preparation}}},
  author = {Bomb{\'i}n, H{\'e}ctor and Pant, Mihir and Roberts, Sam and Seetharam, Karthik I.},
  year = 2024,
  journal = {PRX Quantum},
  volume = {5},
  number = {1},
  pages = {010302},
  doi = {10.1103/PRXQuantum.5.010302}
}

@article{mitigating_smith_2024,
  title = {Mitigating Errors in Logical Qubits},
  author = {Smith, Samuel C. and Brown, Benjamin J. and Bartlett, Stephen D.},
  year = {2024},
  journal = {Commun. Phys.},
  volume = {7},
  number = {1},
  pages = {386},
  doi = {10.1038/s42005-024-01883-4}
}

@article{magic_litinski_2019,
  title = {Magic {{State Distillation}}: {{Not}} as {{Costly}} as {{You Think}}},
  shorttitle = {Magic {{State Distillation}}},
  author = {Litinski, Daniel},
  year = {2019},
  journal = {Quantum},
  volume = {3},
  pages = {205},
  doi = {10.22331/q-2019-12-02-205},
}

\end{document}


\title{Supplementary Information for \\``A superconducting surface-code processor with lattice-surgery logical operations''}

\newcommand{\zju}{School of Physics, ZJU-Hangzhou Global Scientific and Technological Innovation Center, \\and Zhejiang Key Laboratory of Micro-nano Quantum Chips and Quantum Control, Zhejiang University, Hangzhou, China}
\newcommand{\gscaep}{Graduate School of China Academy of Engineering Physics, Beijing, China}

\author{Yanzhe Wang}
\thanks{These authors contributed equally.}
\affiliation{\zju}

\author{Fanhao Shen}
\thanks{These authors contributed equally.}
\affiliation{\zju}

\author{Haipeng Xie}
\affiliation{\gscaep}

\author{Aosai Zhang} 
\affiliation{\zju}
\author{Yu Gao}
\affiliation{\zju}

\author{Chuanyu Zhang}
\affiliation{\zju}
\author{Xuhao Zhu}
\affiliation{\zju}
\author{Feitong Jin}
\affiliation{\zju}
\author{Yiren Zou} 
\affiliation{\zju}
\author{Ning Wang}
\affiliation{\zju}
\author{Zhengyi Cui}
\affiliation{\zju}
\author{Zehang Bao}
\affiliation{\zju}
\author{Zitian Zhu}
\affiliation{\zju}
\author{Jiarun Zhong}
\affiliation{\zju}
\author{Gongyu Liu}
\affiliation{\zju}
\author{Jia-Nan Yang}
\affiliation{\zju}
\author{Yihang Han}
\affiliation{\zju}
\author{Yiyang He}
\affiliation{\zju}
\author{Jiayuan Shen}
\affiliation{\zju}
\author{Han Wang}
\affiliation{\zju}
\author{Jiahua Huang}
\affiliation{\zju}
\author{Xinrong Zhang}
\affiliation{\zju}
\author{Sailang Zhou}
\affiliation{\zju}

\author{Hang Dong}
\affiliation{\zju}
\author{Jinfeng Deng}
\affiliation{\zju}
\author{Yaozu Wu}
\affiliation{\zju}
\author{Zixuan Song}
\affiliation{\zju}

\author{Hekang Li}
\affiliation{\zju}
\author{Zhen Wang}
\affiliation{\zju}
\author{Chao Song}
\affiliation{\zju}
\author{Qiujiang Guo}
\affiliation{\zju}

\author{Pengfei Zhang}
\email{pfzhang@zju.edu.cn}
\affiliation{\zju}

\author{H. Wang}
\email{hhwang@zju.edu.cn}
\affiliation{\zju}

\author{Ying Li}
\email{yli@gscaep.ac.cn}
\affiliation{\gscaep}

\maketitle

\tableofcontents

\section{Device characterization}

\begin{figure*}[tbp]
\centering
\includegraphics[width=\textwidth]{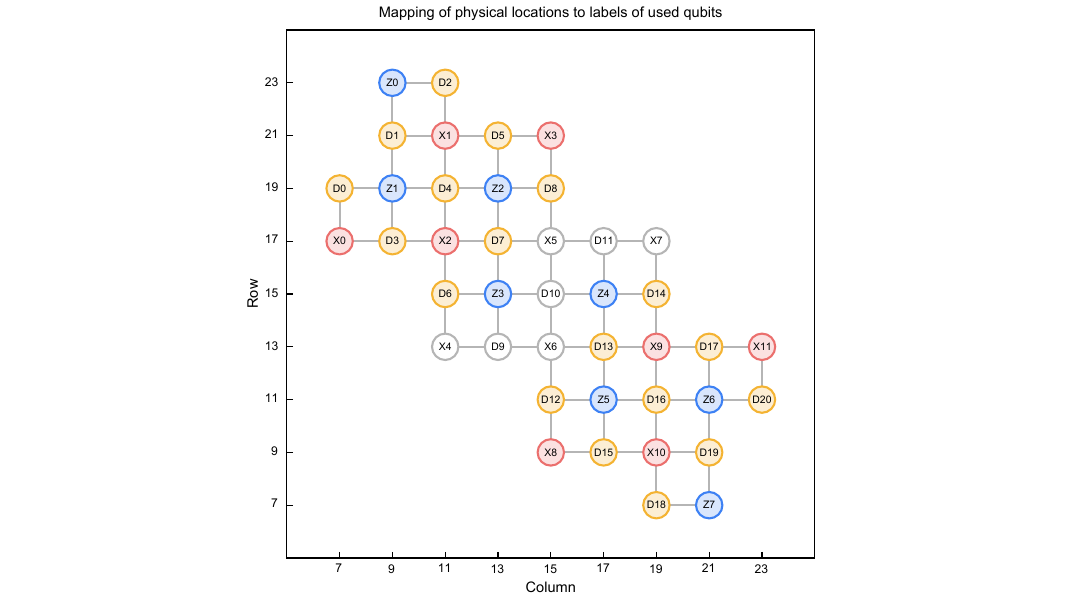}
\caption{
\textbf{Mapping between qubit labels used in Fig.~1 and the main text, and their physical locations shown in the following figures.} 
}
\label{sup_fig_mapping_loc_to_qlabel}
\end{figure*}

\begin{figure*}[tbp]
\centering
\includegraphics[width=\textwidth]{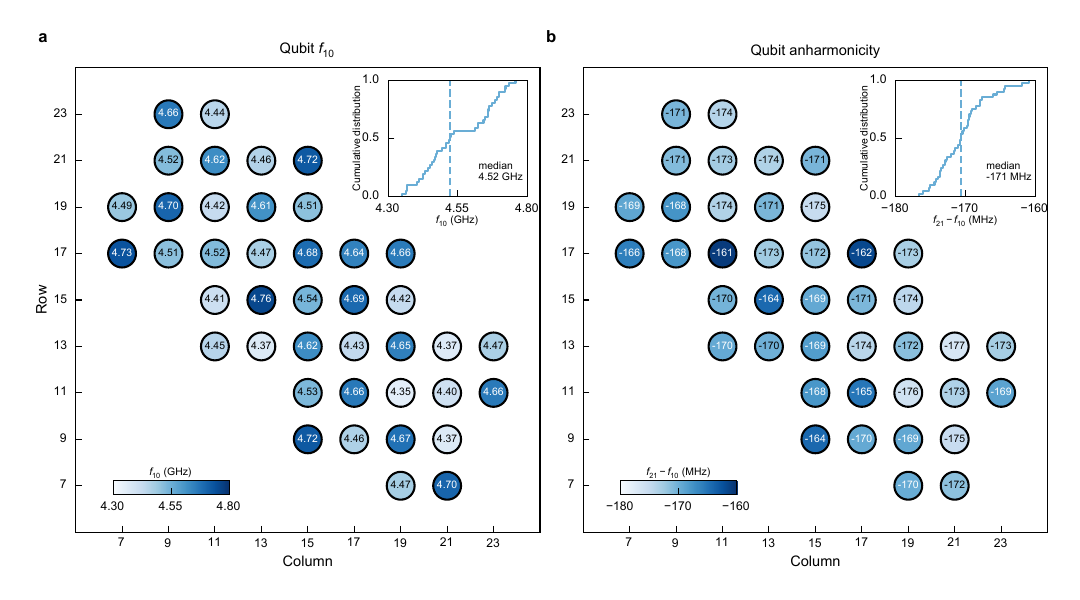}
\caption{
\textbf{Qubit frequency configuration.} 
\textbf{a}, Qubit idle frequency. Inset shows the cumulative distribution of idle frequencies, with the dashed line indicating the median value $4.52$~GHz. 
\textbf{b}, Qubit anharmonicity. The median anharmonicity is $-171$~MHz.
}
\label{sup_fig_f10_and_alpha}
\end{figure*}

\begin{figure*}[tbp]
\centering
\includegraphics[width=\textwidth]{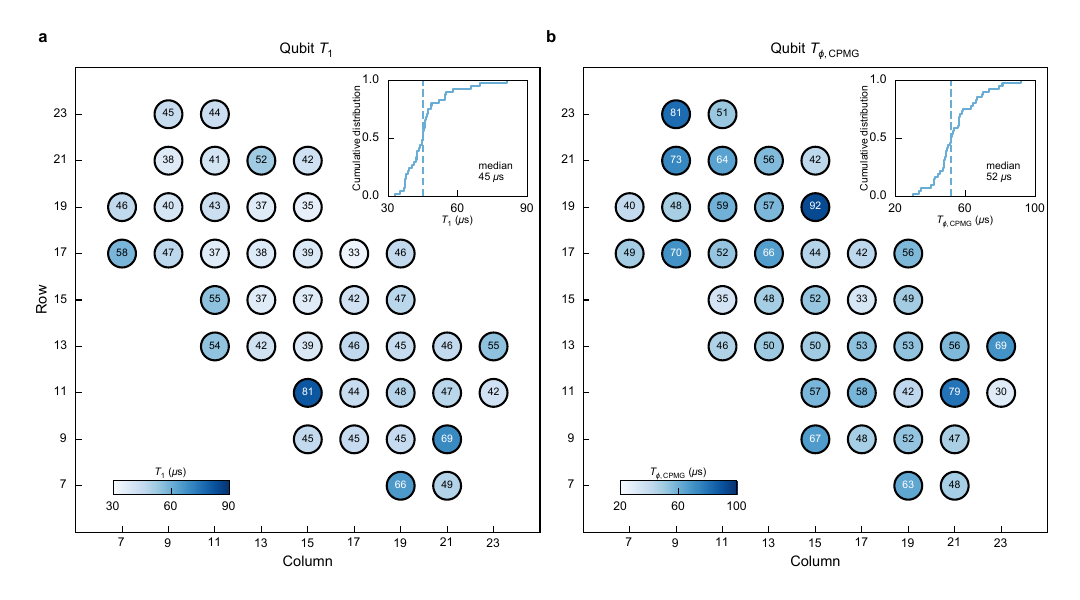}
\caption{
\textbf{Qubit coherence times.} 
\textbf{a}, Energy relaxation time. For each qubit, the energy relaxation time $T_1$ is measured at idle frequency. Inset shows the cumulative distribution, with a median value of $45$~$\mu$s.
\textbf{b}, Pure dephasing time. The pure dephasing time $T_{\phi,\mathrm{CPMG}}$ is measured at the idle frequency using a CPMG sequence. The median value is $52$~$\mu$s.
}
\label{sup_fig_T1s_and_T2s}
\end{figure*}

\begin{figure*}[tbp]
\centering
\includegraphics[width=\textwidth]{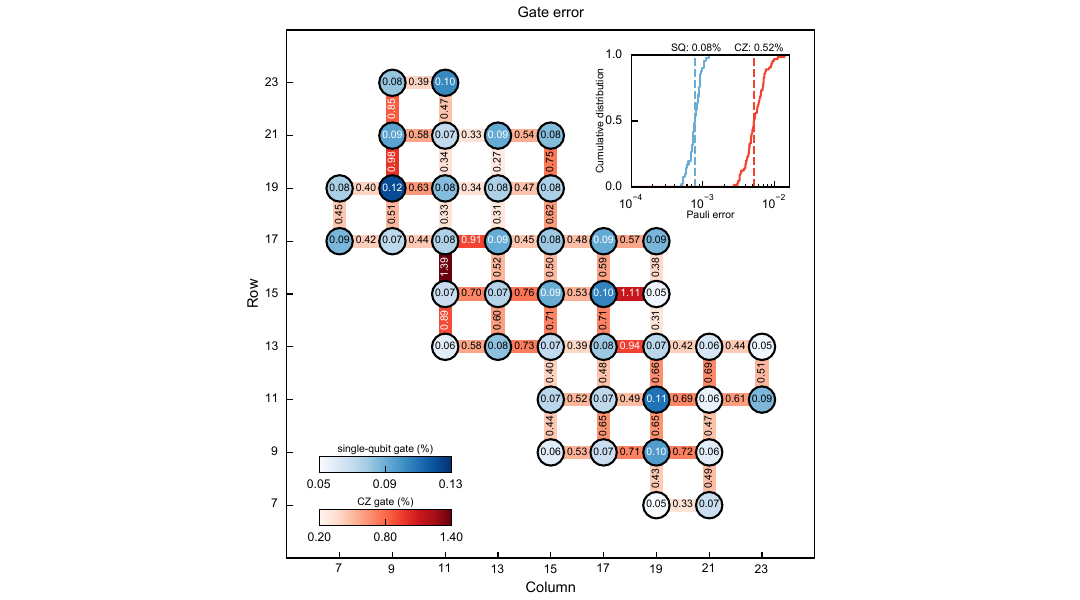}
\caption{
\textbf{Gate errors of single-qubit and CZ gates.} 
Inset shows the cumulative distributions. The median Pauli errors of the single-qubit and CZ gates are $0.08\%$ and $0.52\%$, respectively.
}
\label{sup_fig_SQ_and_CZ}
\end{figure*}

Our quantum processor is a two-dimensional superconducting flip-chip device with nearest-neighbor couplings that enable high-fidelity controlled-$\pi$ phase (CZ) gates, as well as Purcell filters that support fast readout within $500$~ns~\cite{Shen2026Nat.Phys.} (Fig.~\ref{sup_fig_mapping_loc_to_qlabel}).
All qubits are biased at their respective idle points during single-qubit gate operations.
As shown in Fig.~\ref{sup_fig_f10_and_alpha}, the median qubit idle frequency is $4.52$~GHz, and the median qubit anharmonicity is $-171$~MHz. The energy relaxation time ($T_1$) and dephasing time~\cite{10.1063/1.1716296} ($T_{\phi,\mathrm{CPMG}}$), both measured at the idle frequencies, have median values of $45$~$\mu$s and $52$~$\mu$s, respectively; see Fig.~\ref{sup_fig_T1s_and_T2s} for details.
Single-qubit gates are implemented using $22$-ns microwave pulses with derivative removal by adiabatic gate (DRAG) shaping~\cite{Motzoi2009Phys.Rev.Lett.}. CZ gates are realized by tuning the $\ket{11}$ and $\ket{20}$ (or $\ket{02}$) states of two nearest-neighbor qubits close to resonance and activating their coupling for $24$~ns; the qubit and coupler frequencies are optimized to maximize gate fidelity.
The performance of single-qubit and CZ gates is characterized by cross entropy benchmarking (XEB)~\cite{arute2019quantum}, yielding median Pauli errors of $0.08\%$ and $0.52\%$, respectively; see Fig.~\ref{sup_fig_SQ_and_CZ}.

\begin{figure*}[tbp]
\centering
\includegraphics[width=\textwidth]{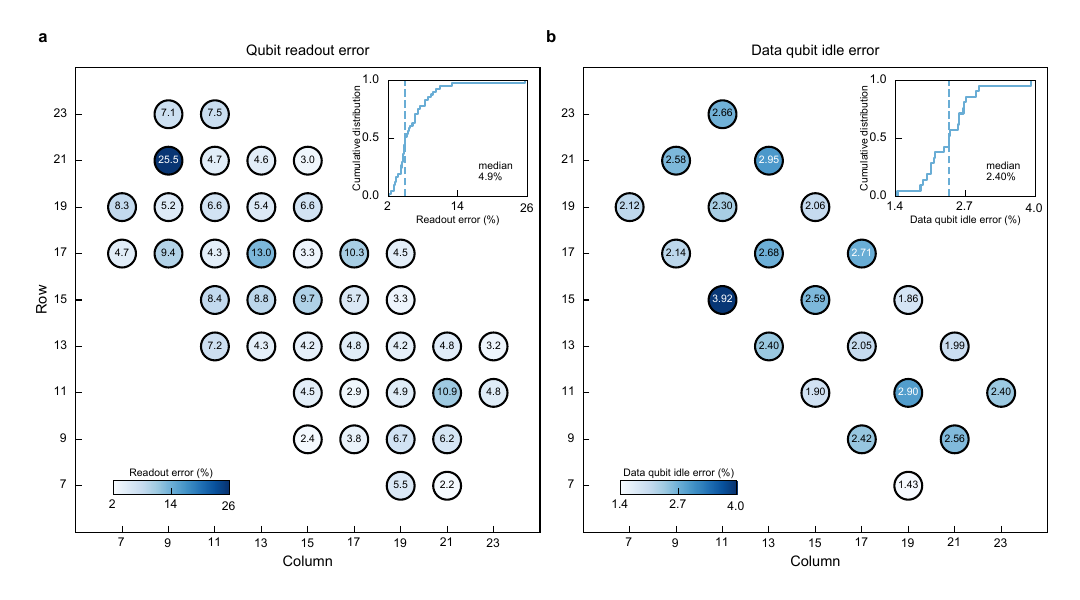}
\caption{
\textbf{Qubit readout error and data-qubit idling error.} 
\textbf{a}, Qubit readout assignment error for three-state discrimination. Inset shows the cumulative distributions, with the dashed line indicating the median value $4.9\%$.
\textbf{b}, Data-qubit idling operation error during the measurement of syndrome qubits. The median value is $2.4\%$.
}
\label{sup_fig_readout_and_DD}
\end{figure*}

In our quantum error correction (QEC) experiments, each syndrome extraction cycle (SEC) consists of the following three steps.
First, interleaved layers of single- and two-qubit gates are applied to the data and syndrome qubits to extract the Z- and X-type stabilizers that define the logical qubit~\cite{GoogleQuantumAI2023Nature}.
Second, the syndrome qubits (or ancilla data qubits) are measured within $500$~ns to determine the stabilizer values, which subsequently serve as the input for decoding algorithms~\cite{Gidney2021Quantum, Higgott2023Phys.Rev.X}. To detect leakage out of the computational subspace, we employ three-state discrimination and discard experimental runs in which any syndrome or data qubit is measured in the state $\ket{2}$. The measurement assignment errors are shown in Fig.~\ref{sup_fig_readout_and_DD}a, with a median value of $4.9\%$.
Third, all qubits are idled for an additional $300$~ns to allow for the depletion of photons in the readout resonators.
To protect the encoded logical information against dephasing noise, we apply Carr-Purcell-Meiboom-Gill (CPMG) dynamical decoupling, implemented as four $X$ gates, to the data qubits during the second and third steps. As shown in Fig.~\ref{sup_fig_readout_and_DD}b, the median Pauli error associated with data-qubit idling is $2.40\%$.

\section{Performance of logical qubit memories}

\begin{figure*}[tbp]
\centering
\includegraphics[width=\textwidth]{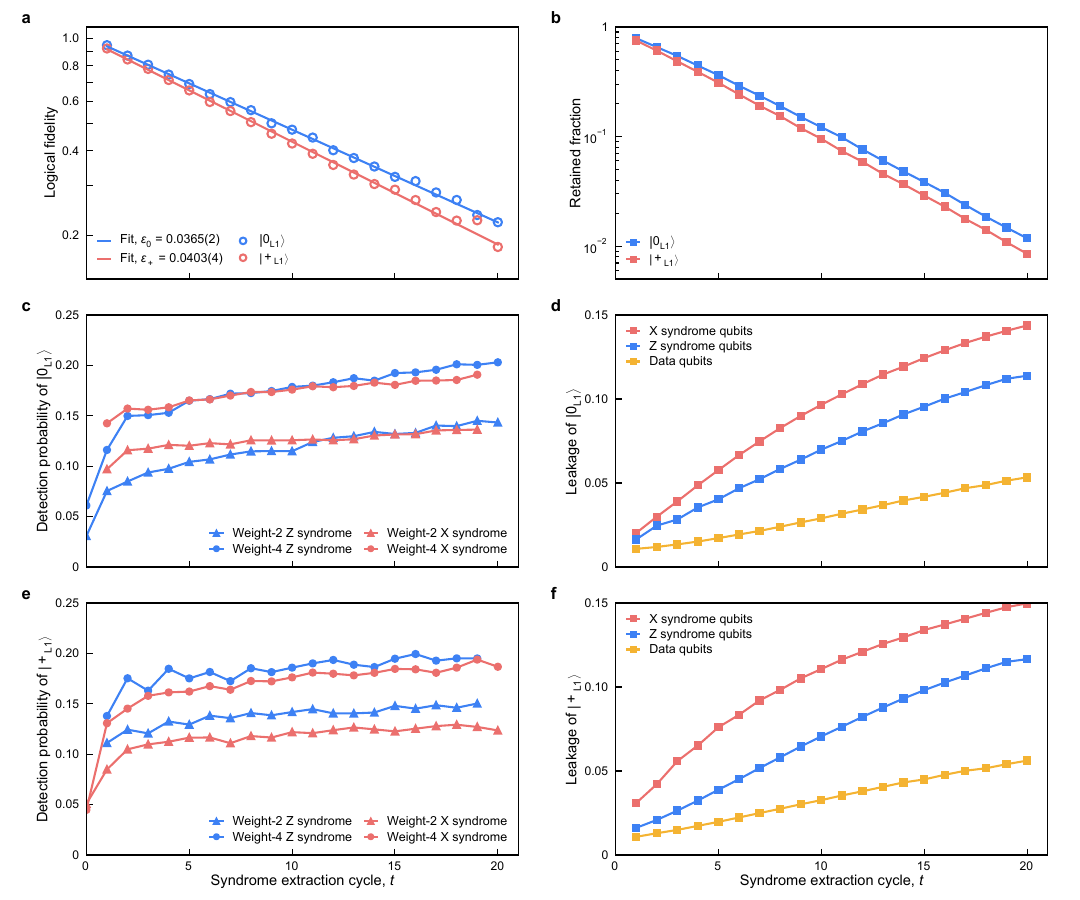}
\caption{
\textbf{Performance of state-preservation experiments for logical qubit L1.}
\textbf{a}, Logical fidelity as a function of SECs. Blue and red circles represent the results of preserving the logical state $\ket{0_\mathrm{L1}}$ and state $\ket{+_\mathrm{L1}}$, respectively. Solid lines represent fits to the experimental data, yielding per-cycle logical error rates $0.0365(2)$ for logical state $\ket{0_\mathrm{L1}}$ and $0.0403(4)$ for logical state $\ket{+_\mathrm{L1}}$.
\textbf{b}, Post-selection efficiency for leakage removal. The plot shows the fraction of experimental runs retained after discarding those with detected leakage events. This fraction decays exponentially, demonstrating the infeasibility of the detection-and-post-selection strategy for large systems with deep error correction cycles. 
\textbf{c}, Averaged detection event probability as a function of SECs for preserving the logical state $\ket{0_\mathrm{L1}}$. Triangles: weight-2 syndrome qubits; circles: weight-4 syndrome qubits; red: X-type syndrome qubits; blue: Z-type syndrome qubits. 
\textbf{d}, Average leakage rate as a function of SECs for preserving the logical state $\ket{0_\mathrm{L1}}$. Red: X-type syndrome qubits; blue: Z-type syndrome qubits; orange: data qubits. 
\textbf{e}, Averaged detection event probability as a function of SECs for preserving the logical state $\ket{+_\mathrm{L1}}$. 
\textbf{f}, Average leakage rate as a function of SECs for preserving the logical state $\ket{+_\mathrm{L1}}$. 
}
\label{sup_figLQL1}
\end{figure*}

\begin{figure*}[tbp]
\centering
\includegraphics[width=\textwidth]{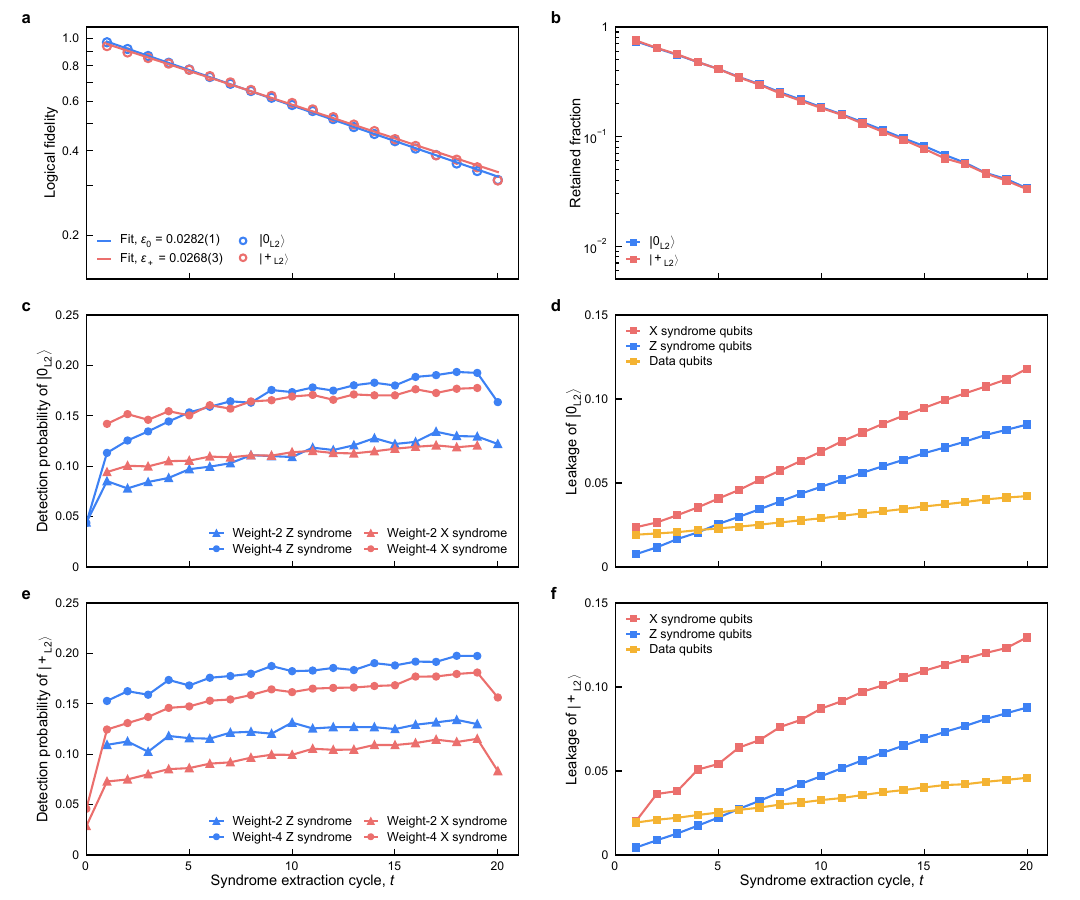}
\caption{
\textbf{Performance of state-preservation experiments for logical qubit L2.}
Fitting the experimental data gives per-cycle logical error rates $0.0282(1)$ for logical state $\ket{0_\mathrm{L2}}$ and $0.0268(3)$ for logical state $\ket{+_\mathrm{L2}}$.
}
\label{sup_figLQL2}
\end{figure*}

In the main text, we present the measured logical fidelities as functions of the number of SECs, with logical qubits L1 and L2 initialized in the states $\ket{0_\mathrm{L1}}$ and $\ket{0_\mathrm{L2}}$, respectively. The logical fidelity is defined as
\begin{equation}
    F = 1 - 2p_\mathrm{L},
\end{equation}
where $p_\mathrm{L}$ denotes the logical error probability. As the number of cycles $t$ increases, the logical fidelity is expected to decay exponentially according to
\begin{equation}
    F = \left(1 - 2\epsilon\right)^{t - t_0},
\end{equation}
where $\epsilon$ is the per-cycle logical error rate and $t_0$ is an offset in the cycle index; both are treated as fitting parameters.

In Figs.~\ref{sup_figLQL1} and~\ref{sup_figLQL2}, we present not only the results for preserving the states $\ket{0_\mathrm{L1}}$ and $\ket{0_\mathrm{L2}}$, but also those for preserving the logical states $\ket{+_\mathrm{L1}}$ and $\ket{+_\mathrm{L2}}$. 
For logical qubit L2, the extracted per-cycle logical error rates show slight differences between these two initial states, which may arise from asymmetry between bit-flip and phase-flip error channels.
In addition, we present the retained fractions, detection probabilities, and leakage populations as functions of the cycle number. As discussed above, we mitigate leakage out of the computational subspace through detection and post-selection. However, this protocol is not scalable because the retained fraction decreases exponentially as the number of error-correction cycles increases. For example, at cycle number $t = 20$, only about $1.2\%$ of the experimental runs remain free of leakage events for L1. 
Because leakage is particularly detrimental to quantum error correction, we carefully arrange the qubit frequencies such that the data qubits have lower frequencies than the syndrome qubits during the CZ gates. This configuration minimizes leakage in the data qubits, while leakage in the syndrome qubits can be reliably detected and subsequently removed through post-selection. 
While this scalability constraint could be further addressed by incorporating active leakage reduction units (LRUs) on both data and syndrome qubits~\cite{McEwen2021Nat.Commun., Marques2023Phys.Rev.Lett.}, we leave such implementations for future work. 
Regarding the detection probabilities, we observe that values associated with weight-2 stabilizers are lower than those for weight-4 stabilizers, as expected from the reduced number of interacting qubits. These probabilities still accumulate with increasing cycle number, indicating residual leakage in the data qubits.

\section{Simplified circuit of two-qubit Deutsch-Jozsa algorithm}

\begin{figure*}[tbp]
\centering
\includegraphics[width=\textwidth]{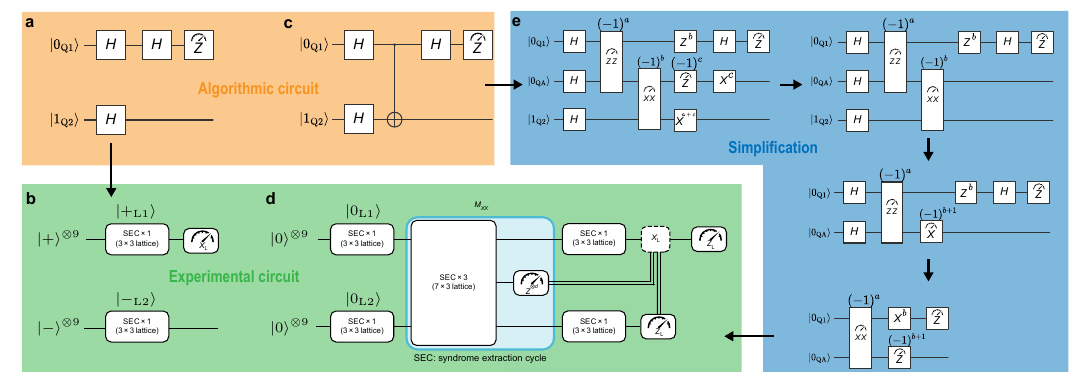}
\caption{
\textbf{Simplified circuit of two-qubit Deutsch-Jozsa algorithm.}
\textbf{a}-\textbf{b}, Algorithmic and experimental circuits for the constant function.
\textbf{c}-\textbf{d}, Algorithmic and experimental circuits for the balanced function. 
\textbf{e,} Simplification procedure from the algorithmic circuit to the experimental circuit for the balanced function.
}
\label{fig:simplification}
\end{figure*}

We experimentally implement fault-tolerant circuits for the Deutsch-Jozsa algorithm corresponding to both constant and balanced functions. As shown in Fig.~\ref{fig:simplification}a, the circuit for the constant function is trivial, as it involves no entanglement between the two qubits $\mathrm{Q}1$ and $\mathrm{Q}2$. In Fig.~\ref{fig:simplification}c, the experimental implementation can be equivalently described as preparing two logical qubits $\mathrm{L}1$ and $\mathrm{L}2$ in the states $\ket{+_{\mathrm{L}1}}$ and $\ket{-_{\mathrm{L}2}}$, respectively, followed by a logical $X$ measurement on the first qubit.

For the balanced function (Fig.~\ref{fig:simplification}b), the algorithm requires a controlled-NOT (CNOT) gate between $\mathrm{Q}1$ and $\mathrm{Q}2$. In a lattice-surgery-based implementation, this CNOT operation necessitates an additional ancilla qubit $\mathrm{QA}$. Below, we describe how this setup can be equivalently reduced to a two-logical-qubit configuration, as shown in Fig.~\ref{fig:simplification}d.

The key simplification arises from the fact that the final measurement is performed only on the control qubit $\mathrm{Q}1$, and its outcome is independent of several operations in the full circuit. Figure~\ref{fig:simplification}e illustrates the complete simplification procedure. First, we remove the operations following the joint measurement $M_{XX}$ on $\mathrm{QA}$ and $\mathrm{Q}2$, since they clearly do not affect the final measurement outcome. Prior to the measurement $M_{XX}$ between $\mathrm{Q}2$ and $\mathrm{QA}$, the qubit $\mathrm{Q}2$ is in the state $|-\rangle$. Therefore, this joint measurement can be equivalently replaced by an $X$ measurement on $\mathrm{QA}$ together with a phase factor of $-1$. At this point, since qubit $\mathrm{Q}2$ does not influence the measurement outcome of $\mathrm{Q}1$ throughout the entire circuit, it can be effectively discarded. Consequently, the three-logical-qubit circuit reduces to a two-qubit circuit. Finally, this reduced circuit can be mapped onto an experimental configuration based on joint $M_{XX}$ measurements.

\section{Experimental protocol of arbitrary rotation gates}

\begin{figure*}[tbp]
\centering
\includegraphics[width=\textwidth]{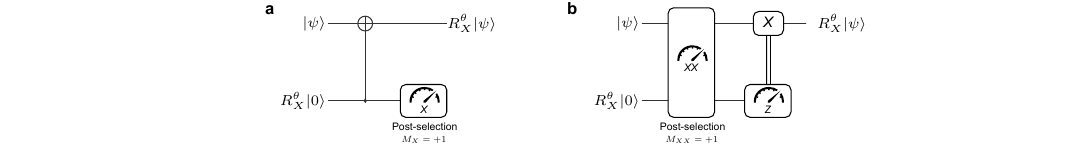}
\caption{
\textbf{Gate teleportation circuit for implementing arbitrary angle rotations.} 
\textbf{a}, Gate teleportation protocol.
\textbf{b}, Replacing the CNOT gate with an $M_{XX}$ measurement. 
}
\label{fig:magic_gate}
\end{figure*}

Gate teleportation is a fundamental technique for implementing non-Clifford gates on encoded quantum states~\cite{Fowler2012Phys.Rev.A}. It is realized by consuming a high-fidelity logical magic state of an ancilla qubit within a fault-tolerant circuit. The fidelity of the magic state can be enhanced in advance using magic-state distillation or cultivation protocols. Figure~\ref{fig:magic_gate}a shows the gate-teleportation circuit for physical qubits, which can be straightforwardly extended to logical qubits by replacing all physical operations with their logical counterparts.
We employ the magic state $\ket{A}=R_X^\theta \ket{0}$ to implement a rotation by angle $\theta$  about the $X$ axis, where $R_X^\theta$ denotes the corresponding unitary operator. For convenience, we express the relevant states in the $\{\ket{+}, \ket{-}\}$ basis. The magic state can then be written as $\ket{A}=\left(\ket{+} + e^{i\theta}\ket{-}\right)\sqrt{2}$.
To teleport the gate operation onto the target qubit, we first apply a CNOT gate to entangle the target qubit with the ancilla. We then perform a destructive $X$-basis measurement $M_X$ on the ancilla and post-select on the outcome $M_X=+1$.
If the target qubit is initially in an arbitrary state $\ket{\psi}=\alpha\ket{+} + \beta \ket{-}$, the final state becomes
\begin{align*}
    \ket{\psi^\prime}
    & = \operatorname{Tr}_\mathrm{A}\left[P_{X, \mathrm{A}}^{+}\, U_{\mathrm{CNOT}}^{\mathrm{A}\to\mathrm{T}}\, \left(\ket{\psi}\otimes\ket{A}\right)\right] \\
    & =\operatorname{Tr}_\mathrm{A}\left[P_{X, \mathrm{A}}^{+}\, U_{\mathrm{CNOT}}^{\mathrm{A}\to\mathrm{T}}\, \left(\alpha\ket{++} + \alpha e^{i\theta}\ket{+-} + \beta\ket{-+} + \beta e^{i\theta}\ket{--}\right)\right] \\
    & =\operatorname{Tr}_\mathrm{A}\left[P_{X, \mathrm{A}}^{+}\, \left(\alpha\ket{++} + \alpha e^{i\theta}\ket{+-} + \beta\ket{--} + \beta e^{i\theta}\ket{-+}\right)\right] \\
    & =\operatorname{Tr}_\mathrm{A}\left[\alpha\ket{++} + \beta e^{i\theta}\ket{-+}\right] \\
    & = \alpha\ket{+} + \beta e^{i\theta}\ket{-} \\
    & = R_X^\theta\ket{\psi},
\end{align*}
where $\operatorname{Tr}_\mathrm{A}[\cdot]$ denotes the partial trace over the ancilla qubit, $P_{X, \mathrm{A}}^{+} = (I+X_\mathrm{A})/2$ is the projector onto the ancilla's $\ket{+}$ state, and $U_{\mathrm{CNOT}}^{\mathrm{A}\to\mathrm{T}}$ is the CNOT unitary acting from ancilla to target.

As shown in Fig.~\ref{fig:magic_gate}b, the CNOT operation can alternatively be replaced using an $M_{XX}$ parity measurement. In this case,
\begin{align*}
    \ket{\psi^\prime}
    & = \operatorname{Tr}_\mathrm{A}\left[P_{Z, \mathrm{A}}^{+}\, P_{XX}^{+}\, \left(\ket{\psi}\otimes\ket{A}\right)\right] \\
    & =\operatorname{Tr}_\mathrm{A}\left[P_{Z, \mathrm{A}}^{+}\, P_{XX}^{+}\, \left(\alpha\ket{++} + \alpha e^{i\theta}\ket{+-} + \beta\ket{-+} + \beta e^{i\theta}\ket{--}\right)\right] \\
    & =\operatorname{Tr}_\mathrm{A}\left[P_{Z, \mathrm{A}}^{+}\, \left(\alpha\ket{++} + \beta e^{i\theta}\ket{--}\right)\right] \\
    & =\operatorname{Tr}_\mathrm{A}\left[\alpha\ket{+0} + \beta e^{i\theta}\ket{-0}\right] \\
    & = \alpha\ket{+} + \beta e^{i\theta}\ket{-} \\
    & = R_X^\theta\ket{\psi},
\end{align*}
where $P_{XX}^{+} = (I+XX)/2$ projects onto the $M_{XX}=+1$ subspace, and 
$P_{Z, \mathrm{A}}^{+} = (I+Z_\mathrm{A})/2$ is the projector onto the ancilla's $\ket{0}$ state.
We post-select runs with measurement outcomes $M_{X}=+1$, $M_{XX}=+1$, and $M_{Z}=+1$; however, post-selection can be removed when real-time feedback is available.
For example, in Fig.~\ref{fig:magic_gate}b, instead of post-selecting the outcome $M_{Z}=+1$, we apply an $X$ correction to the target qubit conditioned on the ancilla's measurement result. This $X$ correction can be further simplified using Pauli-frame tracking, which updates the Pauli operators as $\{X, Y, Z\} \to \{X, -Y, -Z\}$.
Similarly, the post-selection on $M_{X}=+1$ and $M_{XX}=+1$ can be replaced by applying a correction rotation $R_{X}^{2\theta}$ to the target qubit if the measurement outcome is $-1$. In the special case of $\theta=\pi/4$, this correction rotation can also be absorbed into the Pauli-frame tracking, yielding the update rule $\{X, Y, Z\} \to \{X, Z, -Y\}$.

It is interesting to note that, in this special case of $\theta=\pi/4$, we can infer the expectation values of both $Z_\mathrm{L}$ and $Y_\mathrm{L}$ fault-tolerantly by performing only logical $Z$-basis measurements.
If the intermediate joint measurement $M_{XX}$ yields $+1$, we use the logical $Z$-basis outcome as a sample for estimating $\langle Z_\mathrm{L} \rangle$; if $M_{XX}$ yields $-1$, we instead interpret the same outcome as a sample for estimating $-\langle Y_\mathrm{L} \rangle$. Because logical $Z$-basis measurements are fault-tolerant, this procedure provides a fault-tolerant estimate of $Y_\mathrm{L}$ in this special setting. Using this protocol improves the post-selected process fidelity from $0.920^{+10}_{-10}$ to $0.943^{+10}_{-9}$.

\bibliography{references}